\documentclass[letterpaper]{article}

\usepackage[utf8]{inputenc}
\usepackage{arxiv}
\usepackage{float}
\usepackage{enumitem}
\usepackage[english]{babel}
\usepackage{amssymb}
\usepackage{amsmath,mathtools}
\usepackage[table]{xcolor}
\usepackage{graphicx}
\usepackage{multirow}
\usepackage{overpic}
\usepackage{psfrag}
\usepackage{bm}
\usepackage{lineno}
\usepackage{url}
\usepackage{epsfig}
\usepackage{marvosym}
\usepackage{verbatim}
\usepackage{subcaption}
\usepackage{caption}
\usepackage{hyperref}
\hypersetup{colorlinks=true,linkcolor=red,citecolor=blue}
\usepackage{ulem}
\usepackage{lmodern}
\usepackage{booktabs} 
\usepackage{moresize} 
\usepackage{siunitx}
\usepackage{mathtools}
\usepackage{tikz}
\usepackage{pgfplots}
\pgfplotsset{compat=1.14}
\usepgfplotslibrary{fillbetween}
\usetikzlibrary{shapes, arrows, arrows.meta, calc, positioning, decorations,
decorations.pathmorphing, decorations.text, spy}
\pgfplotsset{translate gnuplot=true}
\usepackage{etex} 								
\usepgfplotslibrary{external} 						
\tikzsetexternalprefix{ext/}							
\usepackage{booktabs}
\usepackage{makecell}
\usepackage{rotating,tabularx}
\usepackage{cite}
\definecolor{blue1}	{RGB}{0,177,234}				
\definecolor{blue2}	{RGB}{76,200,239}				
\definecolor{blue3}	{RGB}{127,215,244}				
\definecolor{blue4}	{RGB}{178,231,248}				
\definecolor{blue5}	{RGB}{198,251,255}				
\definecolor{bluegray1}{RGB}{0,127,167}				
\definecolor{bluegray2}{RGB}{76,165,193}				
\definecolor{bluegray3}{RGB}{127,191,211}				
\definecolor{bluegray4}{RGB}{178,216,228}				
\definecolor{gray1}	{RGB}{76,84,93}				
\definecolor{gray2}	{RGB}{129,135,141}				
\definecolor{gray3}	{RGB}{165,169,174}				
\definecolor{gray4}	{RGB}{201,203,206}				
\definecolor{orange1}	{RGB}{255,126,46}				
\definecolor{orange2}	{RGB}{255,164,108}				
\definecolor{orange3}	{RGB}{255,190,150}				
\definecolor{orange4}	{RGB}{255,216,192}				

\newcommand\red[1]{\textcolor{black}{#1}}
\newcommand\blue[1]{\textcolor{black}{#1}}
\newcommand\green[1]{\textcolor{black}{#1}}

\pgfplotsset{
    colormap={custom_map}{[5pt]
            rgb255(0pt)=(255,126,46);
            rgb255(500pt)=(255,190,150);
            rgb255(1000pt)=(0,177,234);
            rgb255(1500pt)=(127,215,244);
    },
}

\newcommand\ie{\textit{i.e.}\,\,}
\newcommand\eg{\textit{e.g.}\,\,}
\newcommand\etal{\textit{et al. \,}}
\renewcommand{\emph}[1]{\textit{#1}}
\setenumerate{label=\textcolor{bluegray1}{$\bm{\diamond}$},itemsep=0.0pt,topsep=5pt}

\renewcommand{\Re}{\operatorname{\mathit{R\kern-.04em e}}} 
\newcommand{\Nu}{\operatorname{\mathit{N\kern-.15em u}}} 
\newcommand{\Ra}{\operatorname{\mathit{R\kern-.04em a}}} 

\title{A review on deep reinforcement learning for fluid mechanics: an update}

\def\size{7.3cm}
\author{
	\parbox{\size}{\centering J. Viquerat\thanks{Corresponding author}}\\
	MINES Paristech, CEMEF\\
	PSL - Research University\\
	\texttt{jonathan.viquerat@mines-paristech.fr}\\
\And
	\parbox{\size}{\centering P. Meliga}\\
	MINES Paristech, CEMEF\\
	PSL - Research University
\And
	\parbox{\size}{\centering A. Larcher}\\
	MINES Paristech, CEMEF\\
	PSL - Research University
\And
	\parbox{\size}{\centering E. Hachem}\\
	MINES Paristech, CEMEF\\
	PSL - Research University
}

\begin{document}
\newgeometry{left=3cm,right=3cm,top=3cm,bottom=2.5cm}
\maketitle

\begin{abstract}
In the past couple of years, the interest of the fluid mechanics community for deep reinforcement learning techniques has increased at fast pace, leading to a growing bibliography on the topic. Due to its ability to solve complex decision-making problems, \blue{deep reinforcement learning} has especially emerged as a valuable tool to perform flow control, but recent publications also advertise great potential for other applications, such as shape optimization or micro-fluidics. The present work proposes an exhaustive review of the existing literature, and is a follow-up to our previous review on the topic. The contributions are regrouped by domain of application, and are compared together regarding algorithmic and technical choices, such as state selection, reward design, time granularity, and more. Based on these comparisons, general conclusions are drawn regarding the current state-of-the-art, and perspectives for future improvements are sketched.
\end{abstract}


\section{Introduction}

During the past decade, machine learning methods, and more specifically deep neural network, have achieved great successes in a wide variety of domains. State-of-the-art neural network architectures have reached astonishing performance levels in image classification tasks \cite{rawat2017, khan2020}, speech recognition \cite{nassif2019} or generative tasks \cite{gui2020}. With a generalized access to GPU computational resources through cheaper hardware or cloud computing, such advances have been paving the way for a general evolution of the reference methods in these domains at both academic and industrial levels.
\red{Machine learning has been making especially rapid inroads in fluid mechanics, as a flexible modeling framework that can be fit to address many challenges, including reduced-order modeling, experimental data processing, shape optimization, turbulence closure modeling, and control~\cite{brunton2020machine}.}

The rapid expansion of neural networks to multiple domains has also yielded important progress in the domain of decision-making techniques, by the coupling of \blue{deep neural networks} with reinforcement learning algorithms (called deep reinforcement learning, or DRL). Several major obstacles that had been hindering classical reinforcement learning have been lifted using the feature extraction capabilities of \blue{deep neural networks} and their ability to handle high-dimensional state spaces. Unprecedented efficiency has been achieved in many domains such as robotics \cite{pinto2017}, language processing \cite{bahdanau2016}, or games \cite{mnih2013, silver2017}, but DRL has also proven useful in many industrial applications, such as autonomous cars \cite{kendall2018, bewley2018}, or data center cooling \cite{googleDataCenter2018}. 

\red{Different incentives trigger the interest of DRL for fluid mechanics applications, that constitutes the core subject of this review:}

\begin{enumerate}
	\item \red{unsteady flow fields exhibiting complex, multiscale phenomena require algorithms capable of handling nonlinearities and multiple spatiotemporal scales. Those are thus particularly amenable to DRL and its representation capabilities, all the more so in the context of flow control, where molding a flow into a more desired state may change the system dynamics and make predictions based on data of uncontrolled systems obsolete, }
	\item \red{in the context of resource expensive numerical environments, DRL offers a way to learn policies by encoding system goals into the reward function and leveraging exploration during training, which is especially beneficial to flow control and optimization problems,}
	\item \red{DRL is a framework that can account for long-term dependencies in decision-making, which is especially interesting to tackle sensitivity to the initial condition or memory effects in turbulence,}
	\item \red{a deeper understanding and exploitation of fluid mechanics is expected to become instrumental in complementing engineering intuition and practical experience, as fluid dynamics has wide applicability, from the way gases circulate around planets, to the way fuel combusts in engines, to the modelling of blood flow and the design of medical implants. }
\end{enumerate}

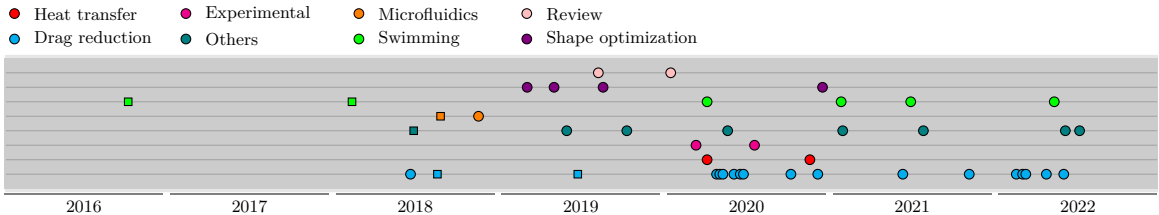
\begin{figure}
\centering
\def\sc{0.73}
\def\papsize{0.176}
\def\linesize{1.5*\papsize}
\def\timelinelength{21}
\def\timelineheight{9*\linesize}
\def\yearlength{2.88}
\def\ctyearlength{3}
\def\yearheight{0.3}
\def\yearbandheight{0.1*\yearheight}
\begin{tikzpicture}	[scale=\sc,
				node distance=0.2cm and 0.1cm,
				every node/.style={	scale=\sc},
				timeline/.style={		rectangle, draw=none, fill=black, opacity=0.2,
								text width=\timelinelength cm, minimum height=\timelineheight cm, text centered,
								inner sep=0pt, outer sep=0pt},
				border/.style={		very thick, draw=black, opacity=0.1},
				year/.style={		text width=\yearlength cm, minimum height=\yearheight cm, text centered, 
								inner sep=0pt, outer sep=0pt, font=\footnotesize},
				yearband/.style={	rectangle, minimum width=\yearlength cm, minimum height=\yearbandheight cm, 
								inner sep=0pt, outer sep=0pt, fill=black, opacity=0.5},
				line/.style={		yearband, minimum width=\timelinelength cm, opacity=0.2},
				lr/.style={			inner sep=0pt, outer sep=0pt},
				paper/.style={		circle, fill=black, draw=black, inner sep=0pt, minimum size=\papsize cm},
				previous/.style={	regular polygon, regular polygon sides=4, fill=black, draw=black, inner sep=0pt, minimum size=1.1*\papsize cm},
				drag/.style={		paper, fill=cyan},
				pdrag/.style={		previous, fill=cyan},
				heat/.style={		paper, fill=red},
				shapeopt/.style={	paper, fill=violet},
				exp/.style={		paper, fill=magenta},
				micro/.style={		paper, fill=orange},
				pmicro/.style={		previous, fill=orange},
				swimming/.style={	paper, fill=green},
				pswimming/.style={	previous, fill=green},
				review/.style={		paper, fill=pink},
				other/.style={		paper, fill=teal},
				pother/.style={		previous, fill=teal},
				legend/.style={		font=\footnotesize}
				]

	\node[timeline] 	(timeline) at (0.5*\timelinelength,0) {} ;
	\draw[border] 	($(timeline.north west) + (0pt,1pt)$) -- ($(timeline.north east) + (0pt,1pt)$);
	\draw[border] 	($(timeline.south west) + (0pt,-1pt)$) -- ($(timeline.south east) + (0pt,-1pt)$);

	\node[year, anchor=north west] (2016) at ($(timeline.south west) + (0pt,-5pt)$) {2016};
	\node[year, right=of 2016] (2017) {2017};
	\node[year, right=of 2017] (2018) {2018};
	\node[year, right=of 2018] (2019) {2019};
	\node[year, right=of 2019] (2020) {2020};
	\node[year, right=of 2020] (2021) {2021};
	\node[year, right=of 2021] (2022) {2022};
	
	\node[yearband] at ($(2016.north) + (0pt,2pt)$) {};
	\node[yearband] at ($(2017.north) + (0pt,2pt)$) {};
	\node[yearband] at ($(2018.north) + (0pt,2pt)$) {};
	\node[yearband] at ($(2019.north) + (0pt,2pt)$) {};
	\node[yearband] at ($(2020.north) + (0pt,2pt)$) {};
	\node[yearband] at ($(2021.north) + (0pt,2pt)$) {};
	\node[yearband] at ($(2022.north) + (0pt,2pt)$) {};
	
	\node[lr] (bot_left) at (timeline.west |- timeline.south) {};
	\node[lr] (l_l1) at ($(bot_left) + (0pt, 1*\linesize cm)$) {};
	\node[lr] (l_l2) at ($(bot_left) + (0pt, 2*\linesize cm)$) {};
	\node[lr] (l_l3) at ($(bot_left) + (0pt, 3*\linesize cm)$) {};
	\node[lr] (l_l4) at ($(bot_left) + (0pt, 4*\linesize cm)$) {};
	\node[lr] (l_l5) at ($(bot_left) + (0pt, 5*\linesize cm)$) {};
	\node[lr] (l_l6) at ($(bot_left) + (0pt, 6*\linesize cm)$) {};
	\node[lr] (l_l7) at ($(bot_left) + (0pt, 7*\linesize cm)$) {};
	\node[lr] (l_l8) at ($(bot_left) + (0pt, 8*\linesize cm)$) {};

	\node[lr] (bot_right) at (timeline.east |- timeline.south) {};
	\node[lr] (r_l1) at ($(bot_right) + (0pt, 1*\linesize cm)$) {};
	\node[lr] (r_l2) at ($(bot_right) + (0pt, 2*\linesize cm)$) {};
	\node[lr] (r_l3) at ($(bot_right) + (0pt, 3*\linesize cm)$) {};
	\node[lr] (r_l4) at ($(bot_right) + (0pt, 4*\linesize cm)$) {};
	\node[lr] (r_l5) at ($(bot_right) + (0pt, 5*\linesize cm)$) {};
	\node[lr] (r_l6) at ($(bot_right) + (0pt, 6*\linesize cm)$) {};
	\node[lr] (r_l7) at ($(bot_right) + (0pt, 7*\linesize cm)$) {};
	\node[lr] (r_l8) at ($(bot_right) + (0pt, 8*\linesize cm)$) {};
	
	\draw[line] (l_l1) -- (r_l1);
	\draw[line] (l_l2) -- (r_l2);
	\draw[line] (l_l3) -- (r_l3);
	\draw[line] (l_l4) -- (r_l4);
	\draw[line] (l_l5) -- (r_l5);
	\draw[line] (l_l6) -- (r_l6);
	\draw[line] (l_l7) -- (r_l7);
	\draw[line] (l_l8) -- (r_l8);
	
	\node (p) at (2018.west |- l_l1) {};
	\node[drag] at ($(p) + (0.47*\yearlength,0 pt)$) {}; 
	\node[pdrag] at ($(p) + (0.64*\yearlength,0 pt)$) {}; 
	\node (p) at (2019.west |- l_l1) {};
	\node[pdrag] at ($(p) + (0.48*\yearlength,0 pt)$) {}; 
	\node (p) at (2020.west |- l_l1) {};
	\node[drag] at ($(p) + (0.31*\yearlength,0 pt)$) {}; 
	\node[drag] at ($(p) + (0.33*\yearlength,0 pt)$) {}; 
	\node[drag] at ($(p) + (0.35*\yearlength,0 pt)$) {}; 
	\node[drag] at ($(p) + (0.42*\yearlength,0 pt)$) {}; 
	\node[drag] at ($(p) + (0.46*\yearlength,0 pt)$) {}; 
	\node[drag] at ($(p) + (0.48*\yearlength,0 pt)$) {}; 
	\node[drag] at ($(p) + (0.78*\yearlength,0 pt)$) {}; 
	\node[drag] at ($(p) + (0.95*\yearlength,0 pt)$) {}; 
	\node (p) at (2021.west |- l_l1) {};
	\node[drag] at ($(p) + (0.44*\yearlength,0 pt)$) {}; 
	\node[drag] at ($(p) + (0.86*\yearlength,0 pt)$) {}; 
	\node (p) at (2022.west |- l_l1) {};
	\node[drag] at ($(p) + (0.11*\yearlength,0 pt)$) {}; 
	\node[drag] at ($(p) + (0.41*\yearlength,0 pt)$) {}; 
	\node[drag] at ($(p) + (0.30*\yearlength,0 pt)$) {}; 
	\node[drag] at ($(p) + (0.15*\yearlength,0 pt)$) {}; 
	\node[drag] at ($(p) + (0.17*\yearlength,0 pt)$) {}; 
	
	\node (p) at (2020.west |- l_l2) {};
	\node[heat] at ($(p) + (0.25*\yearlength,0 pt)$) {}; 
	\node[heat] at ($(p) + (0.90*\yearlength,0 pt)$) {}; 
		
	\node (p) at (2020.west |- l_l3) {};
	\node[exp] at ($(p) + (0.18*\yearlength,0 pt)$) {}; 
	\node[exp] at ($(p) + (0.55*\yearlength,0 pt)$) {}; 
	
	\node (p) at (2018.west |- l_l4) {};
	\node[pother] at ($(p) + (0.49*\yearlength,0 pt)$) {}; 
	\node (p) at (2019.west |- l_l4) {};
	\node[other] at ($(p) + (0.41*\yearlength,0 pt)$) {}; 
	\node[other] at ($(p) + (0.79*\yearlength,0 pt)$) {}; 
	\node (p) at (2020.west |- l_l4) {};
	\node[other] at ($(p) + (0.38*\yearlength,0 pt)$) {}; 
	\node (p) at (2021.west |- l_l4) {};
	\node[other] at ($(p) + (0.06*\yearlength,0 pt)$) {}; 
	\node[other] at ($(p) + (0.57*\yearlength,0 pt)$) {}; 
	\node (p) at (2022.west |- l_l4) {};
	\node[other] at ($(p) + (0.42*\yearlength,0 pt)$) {}; 
	\node[other] at ($(p) + (0.51*\yearlength,0 pt)$) {}; 
	
	\node (p) at (2018.west |- l_l5) {};
	\node[pmicro] at ($(p) + (0.66*\yearlength,0 pt)$) {}; 
	\node[micro] at ($(p) + (0.90*\yearlength,0 pt)$) {}; 
	
	\node (p) at (2016.west |- l_l6) {};
	\node[pswimming] at ($(p) + (0.78*\yearlength,0 pt)$) {}; 
	\node (p) at (2018.west |- l_l6) {};
	\node[pswimming] at ($(p) + (0.10*\yearlength,0 pt)$) {}; 
	\node (p) at (2020.west |- l_l6) {};
	\node[swimming] at ($(p) + (0.25*\yearlength,0 pt)$) {}; 
	\node (p) at (2021.west |- l_l6) {};
	\node[swimming] at ($(p) + (0.05*\yearlength,0 pt)$) {}; 
	\node[swimming] at ($(p) + (0.49*\yearlength,0 pt)$) {}; 
	\node (p) at (2022.west |- l_l6) {};
	\node[swimming] at ($(p) + (0.35*\yearlength,0 pt)$) {}; 
	
	\node (p) at (2019.west |- l_l7) {};
	\node[shapeopt] at ($(p) + (0.16*\yearlength,0 pt)$) {}; 
	\node[shapeopt] at ($(p) + (0.64*\yearlength,0 pt)$) {}; 
	\node (p) at (2019.west |- l_l7) {};
	\node[shapeopt] at ($(p) + (0.33*\yearlength,0 pt)$) {}; 
	\node (p) at (2020.west |- l_l7) {};
	\node[shapeopt] at ($(p) + (0.98*\yearlength,0 pt)$) {}; 
	
	\node (p) at (2019.west |- l_l8) {};
	\node[review] at ($(p) + (0.61*\yearlength,0 pt)$) {}; 
	\node (p) at (2020.west |- l_l8) {};
	\node[review] at ($(p) + (0.02*\yearlength,0 pt)$) {}; 
	
	\node[drag] (drag_legend) at ($(timeline.north west) + (5pt, 10 pt)$) {};
	\node[heat, above=of drag_legend] (heat_legend) {};
	
	\node[legend, right=of drag_legend] (drag_legend_txt) {Drag reduction};
	\node[legend, right=of heat_legend] (heat_legend_txt) {Heat transfer};
	
	\node[exp, right=of heat_legend_txt, xshift=15pt] (exp_legend) {};
	\node[other, below=of exp_legend] (other_legend) {};
	
	\node[legend, right=of exp_legend] (exp_legend_txt) {Experimental};
	\node[legend, right=of other_legend] (other_legend_txt) {Others};
	
	\node[micro, right=of exp_legend_txt, xshift=15pt] (micro_legend) {};
	\node[swimming, below=of micro_legend] (swimming_legend) {};
	
	\node[legend, right=of micro_legend] (micro_legend_txt) {Microfluidics};
	\node[legend, right=of swimming_legend] (swimming_legend_txt) {Swimming};
	
	\node[review, right=of micro_legend_txt, xshift=15pt] (review_legend) {};
	\node[shapeopt, below=of review_legend] (shapeopt_legend) {};
	
	\node[legend, right=of review_legend] (review_legend_txt) {Review};
	\node[legend, right=of shapeopt_legend] (shapeopt_legend_txt) {Shape optimization};

\end{tikzpicture}
\caption{\textbf{Timeline of recent publications on deep reinforcement learning for fluid dynamics.} Colors indicate different fields of application. Please note that we retain here the date of the first pre-print publication, and not that of final publication in peer-review journals. Indeed, the fast-paced evolution of the \blue{deep reinforcement learning} community brings particular importance to pre-prints, sometimes supported by code release. The square symbols denote references included in our previous review \cite{garnier2021}.} 
\label{fig:degenerate_DRL}
\end{figure}

\red{Despite these motivations,} the efforts for applying DRL to fluid mechanics
are ongoing but still at an early stage, with only a handful of pioneering studies providing insight into the performance improvements to be delivered in the field. Nonetheless, from a few liminal contributions in 2016 \cite{novati2017} and early 2018 \cite{verma2018}, the domain has undergone an increasing inflow of contributions, that pinnacled in 2020, with no less than 16 pre-prints and articles, and a clear focus on drag reduction problems, as shown in figure \ref{fig:degenerate_DRL}. This enthusiasm can be explained by two main factors: the increasing number of open-source initiatives \cite{rabault2019, belus2019, viquerat2021}, that has led to an accelerated diffusion of the methods in the community, and the sustained commitment from the machine learning community, that has allowed concurrently expanding the scope from computationally inexpensive, low-dimensional reductions of the underlying fluid dynamics to complex Navier--Stokes systems, all the way to experimental set-ups.

The present review proposes a six-year perspective of deep reinforcement learning applied to fluid flow problems, in the context of both numerical and experimental environments. It is intended as a follow up to our first review released as a pre-print in 2019 \cite{garnier2021}, that was followed in 2020 by a short review from another group of authors, focused on drag reduction and shape optimization problems \cite{rabault2020}. To the best of the authors knowledge, those are the only other similar initiatives preceding this one, that are also featured in figure \ref{fig:degenerate_DRL} for the sake of completeness. \red{in practice, only contributions focusing  on the application of deep reinforcement learning techniques (not machine learning in a broader sense) to fluid dynamical systems, either experimental or numerical as governed by Navier--Stokes equations (or a reduced model thereof), have been considered. The objective is two-fold: first to analyze the main trends, achievements and research directions currently pursued in the community. Second, to identify the main needs to inform future developments towards practical deployment.}

The organization is as follows: a reminder on the main DRL algorithms that have been used in a fluid dynamics context is provided in section \ref{section:background}. Section \ref{section:challenges} lists of relevant issues to consider when evaluating the progress of DRL for practically fluid flow problems. An extensive bibliographical review is then conducted in sections \ref{section:revcfd} and \ref{section:revexpe}, that covers a total of \green{43} papers, most of them subsequent to the previous review articles. Contributions are grouped and compared by domain of applications. Finally, section \ref{section:transverse} draws general conclusions on the state-of-the-art and proposes a transversal study including suggestions for future work in the field.

\section{Deep reinforcement learning}\label{section:background}

This section covers the basic concepts of deep reinforcement learning, and briefly describes the methods most represented in the selected contributions. First, the basic concepts of reinforcement learning are introduced, whereafter value-based and policy-based methods are distinguished. Then, specificities of DRL are detailed, and a curated list of algorithms is proposed, including deep Q-networks (DQN), advantage actor-critic (A2C), proximal policy optimization (PPO), trust-region policy optimization (TRPO), deep deterministic policy gradients (DDPG), twin-delayed deep deterministic policy gradients (TD3), soft actor-critic (SAC) and policy-based optimization (PBO). For a more sophisticated introduction to DRL, the reader is referred to \cite{sutton2018}. The notations used in the remaining of this review can be found in table \ref{table:notations}.

\begin{table}
    \footnotesize
    \caption{\textbf{List of notations and acronysms.}}
    \label{table:notations}
    \centering
    \begin{tabular}{rl}
        \toprule
        $\gamma$ & discount factor\\
        $\lambda$ & learning rate\\
        $\alpha$, $\beta$ & step-size\\
        $\epsilon$ & probability of random action\\\midrule
        $s$,$s'$ & states\\
        $\mathcal{S}^{+}$ & set of all states\\
        $\mathcal{S}$ & set of non-termination states\\
        $a$ & action\\
        $\mathcal{A}$ & set of all actions\\
        $r$ & reward\\
        $\mathcal{R}$ & set of all rewards\\
        $\mathcal{D}$ & set of collected transitions\\
        $t$ & time station\\
        $T$ & final time station\\
        $a_t$ & action at time $t$\\
        $s_t$ & state at time $t$\\
        $r_t$ & reward at time $t$\\
        $r(s,a)$ & reward received for taking action $a$ in state $s$\\
        $R(\tau)$ & discounted cumulative reward following trajectory $\tau$\\
        $G_t$ & discounted cumulative reward starting from time $t$\\\midrule
        $\pi$ & policy\\
        $\theta$, $\theta'$ & parameterization vector of a policy\\
        $\pi_\theta$ & policy parameterized by $\theta$\\
        $\pi (s)$ & action probability distribution in state $s$ following $\pi$\\
        $\pi (a \vert s)$ & probability of taking action $a$ in state $s$ following $\pi$\\
        $V^{\pi}(s)$ & value of state $s$ under policy $\pi$\\
        $V^{*}(s)$ & value of state $s$ under the optimal policy\\
        $Q^{\pi}(s,a)$ & value of taking action $a$ in state $s$ under policy $\pi$\\
        $Q^{*}(s,a)$ & value of taking action $a$ in state $s$ under the optimal policy\\
        $Q_\theta(s,a)$ & estimated value of taking action $a$ in state $s$ with parameterization $\theta$\\\midrule
        \blue{A2C} & \blue{Advantage actor-critic}\\
        \blue{BO} & \blue{Bayesian optimization}\\     
        \blue{CFD} & \blue{Computational fluid dynamics}\\        
        \blue{CMA-ES} & \blue{Covariance matrix adaptation evolution strategy }\\                     
        \blue{DDPG} & \blue{Deep deterministic policy gradient}\\                
        \blue{DQN} & \blue{Deep Q-networks}\\
        \blue{DDQN} & \blue{Double deep Q-networks}\\
        \blue{DRL} & \blue{Deep reinforcement learning}\\                
        \blue{GP} & \blue{Genetic programming}\\
        \blue{LIPO} & \blue{Lipschitz global optimization}\\        
        \blue{LSTM} & \blue{Long-short term memory}\\                        
        \blue{MFEC} & \blue{Model-free episodic control}\\           
        \blue{PBO} & \blue{Policy-based optimization}\\        
        \blue{PPO} & \blue{Proximal policy optimization}\\        
        \blue{PPO-1} & \blue{Single-step proximal policy optimization}\\
        \blue{RL} & \blue{Reinforcement learning}\\
        \blue{SAC} & \blue{Soft actor-critic}\\
        \blue{TD3} & \blue{Twin-delayed deep deterministic policy gradient}\\
        \blue{TRPO} & \blue{Trust-region policy optimization}\\
        \bottomrule
    \end{tabular}
\end{table}

\subsection{Reinforcement learning}

\begin{figure}
\centering
\begin{tikzpicture}[	fontsize/.style={		font=\footnotesize},
				intnode/.style={		black, fontsize, pos=0.5},
				arrow/.style={		thick,color=bluegray3, rounded corners,thick,-stealth},
				box/.style={		rectangle,rounded corners, draw=gray1, very thick, 
								text width=2.5cm,minimum height=1.2cm,text centered,
								inner sep=2pt, outer sep=0pt, fontsize}]

	\node[box,fill=blue4] 	(env) 	at (0,0) {Environment\\ $s_t \mapsto s_{t+1}$};
	\node[box,fill=orange4] (agent) 	at (5,0) {Agent};
	
	\draw[arrow] 	(env.east) 		to [out=0,in=180] 	node[intnode, above] {$r_t$} 	(agent.west);
	\draw[arrow] 	(agent.south) 	to [out=-90,in=-90] 	node[intnode, above] {$a_t$} 	(env.south);
	\draw[arrow] 	(env.north) 	to [out=90,in=90] 	node[intnode, below] {$w_{t}$} 	(agent.north);

\end{tikzpicture}
\caption{\textbf{\blue{Reinforcement learning} agent and its interactions with the environment.}}
\label{fig:rl}
\end{figure}
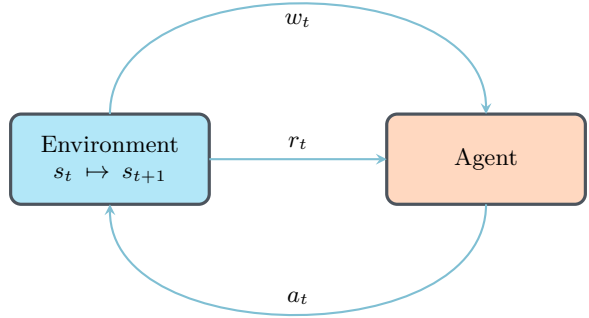

Reinforcement learning \blue{(RL)} is a class of methods designed for decision-making problems, in which an agent learns to interact with an environment by (i) observing it, (ii) taking actions based on these observations, and (iii) receiving rewards from it, as a measure of the quality of the action taken. RL is based on Markov decision processes, for which a typical execution goes as follows (see also figure \ref{fig:rl}):

\begin{enumerate}
	\item Assume the environment is in state $s_{t} \in \mathcal{S}$ at iteration $t$, where $\mathcal{S}$ is a set of states;
	\item The agent uses $w_{t}$, an observation of the current environment state (and possibly a partial subset of $s_t$) to take action $a_{t} \in \mathcal{A}$, where $\mathcal{A}$ is a set of actions;
	\item The environment reacts to the action by transitionning from $s_t$ to state $s_{t+1}\in \mathcal{S}$;
	\item The agent is fed with a reward $r_{t} \in \mathcal{R}$, where $\mathcal{R}$ is a set of rewards, and a new observation $w_{t+1}$.
\end{enumerate}

The steps described above repeat until a termination state is reached, and the succession of states and actions then define a finite trajectory $\tau = \big( s_0, a_0, s_1, a_1, ... \big)$. In any given state, the objective of the agent is to determine the adequate action to maximize its cumulative reward over an episode, \textit{i.e.} over one trajectory. Most often, the quantity of interest is the discounted cumulative reward along a trajectory, defined as:

\begin{equation}
\label{eq:discounted_cumulative_reward}
	R(\tau) = \displaystyle \sum_{t=0}^{T} \gamma^t r_{t}\,,
\end{equation}

where $T$ is the horizon of the trajectory (\textit{i.e.} the terminal time station), and $\gamma\in[0,1]$ is a discount factor that weights the relative importance of present and future rewards. Within the zoology of DRL methods, we distinguish two categories, namely \emph{model-based} and \emph{model-free} algorithms. Model-based method incorporates a model of the environment they interact with, and will not be considered in this paper (the reader is referred to \cite{sutton2018} and references therein for details about model-based methods). On the contrary, model-free algorithms directly interact with their environment, and are currently the most commonly used within the DRL community, mainly for their ease of application and implementation. Model-free methods are further distinguished between \emph{value-based} methods and \emph{policy-based} methods \cite{sutton2018}.  Although both approaches aim at maximizing their  expected return, policy-based methods do so by directly optimizing the parameterized policy, while value-based methods learn to estimate the expected value of a state-action pair optimally, which in turn determines the best action to take in each state. 

\subsubsection{Value-based methods}

In value-based methods, the agent learns to optimally estimate a \emph{value function}, which in turn dictates the policy of the agent by selecting the action of the highest value. One usually defines the \emph{state value function}:

\begin{equation*}
    V^\pi (s) = \underset{\tau \sim \pi}{\mathbb{E}} \big[ R(\tau) \vert s \big],
\end{equation*}

denoting the expected discounted cumulative reward starting in state $s$, then following trajectory $\tau$ according to policy $\pi$, and the \emph{state-action value function}, or Q-function:

\begin{equation*}
    Q^\pi (s,a) = \underset{\tau \sim \pi}{\mathbb{E}} \big[ R(\tau) \vert s, a \big],
\end{equation*}

denoting the same expected discounted cumulative reward starting in state $s$ and taking action $a$. Both values are quite obviously such that:

\begin{equation*}
    V^\pi (s) = \underset{a \sim \pi}{\mathbb{E}} \big[ Q^\pi (s, a) \big],
\end{equation*}

meaning that in practice, $V^\pi (s)$ is the weighted average of $Q^\pi (s, a)$ over all possible actions by the probability of each action. One of the main value-based methods in use is called Q-learning, as it relies on the learning of the Q-function to find an optimal policy. In classical Q-learning, the Q-function is stored in a Q-table, which is a simple array representing the estimated value of the optimal Q-function $Q^*(s, a)$ for each pair $(s,a) \in \mathcal{S} \times \mathcal{A}$. The Q-table is initialized randomly, and its values are progressively updated as the agent explores the environment, until the Bellman optimality condition \cite{bellman1962} is reached: 

\begin{equation}
\label{eq:bellman}
    Q^* (s,a) = r(s,a) + \gamma \max_{a'} Q^* (s',a'),
\end{equation}

at which point the Q-table estimate of the Q-value has converged, and taking the action with the highest Q-value systematically leads to the optimal policy.

\subsubsection{Policy-based methods}

Policy methods maximize the expected discounted cumulative reward of a policy $\pi(a \vert s)$ mapping states to actions, and resort not to a value function, but to a probability distribution over actions given states.
Compared to value-based methods, policy-based methods offer three main advantages:

\begin{enumerate}
	\item they have better convergence properties, although they tend to get trapped in local minima;
	\item they naturally handle high dimensional action spaces;
	\item they can learn stochastic policies.
\end{enumerate}

Most reinforcement learning algorithms applied to fluid mechanics problems are policy gradient methods, in which gradient ascent is
used to optimize a parameterized policy $\pi_\theta(a \vert s)$ with respect to some measure of the expected return.
In practice, one defines an objective function based on the expected discounted cumulative reward:

\begin{equation*}
    J(\theta) = \underset{\tau \sim \pi_\theta}{\mathbb{E}} \big[ R(\tau) \big],
\end{equation*}

and seeks the optimal parameterization $\theta^*$ that maximizes $J(\theta)$:

\begin{equation*}
    \theta^* = \arg \max_\theta \underset{\tau \sim \pi_\theta}{\mathbb{E}} \big[ R(\tau) \big],
\end{equation*}

which can be done on paper by plugging an estimator of the policy gradient $\nabla_\theta J(\theta)$ into a gradient ascent algorithm. In practice, this is no small task, as one is looking for the gradient with respect to the policy parameters $\theta$, in a context where the effects of policy changes on the state distribution are unknown (since modifying the policy will most likely modify the set of visited states, which will in turn affect performance in some indefinite manner). The standard derivation relies on the log-probability trick \cite{williams1992}, and allows expressing $\nabla_\theta J(\theta)$ as an evaluable expected value:

\begin{equation}
\label{eq:policy_gradient}
	\nabla_\theta J(\theta) = \underset{\tau \sim \pi_\theta}{\mathbb{E}} \left[ \sum_{t=0}^T \nabla_\theta \log \left( \pi_\theta (a_t \vert s_t) \right) R(\tau) \right],
\end{equation}

after which the gradient is used to update the policy parameters:

\begin{equation}
\label{eq:theta_update}
	\theta \leftarrow \theta + \lambda \nabla_\theta J(\theta).
\end{equation}

It must be noted that expression (\ref{eq:policy_gradient}) represents the simplest form of the policy gradient, and that, in practice, more elaborate expressions are chosen instead of $R(\tau)$ that decrease the variance of the associated estimator without introducing bias, such as the reward-to-go or the advantage function, among others (see section \ref{section:a2c}).

\subsection{Deep reinforcement learning}

\begin{figure}
\centering
\begin{tikzpicture}[	arrow/.style=		{thick,color=bluegray1,rounded corners},
				netnode/.style=		{circle, inner sep=0pt, text width=22pt, align=center, very thick},
				inputnode/.style=	{netnode, fill=blue4, draw=black},
				hiddennode/.style=	{netnode, fill=gray4, draw=black},
				outputnode/.style=	{netnode, fill=orange4, 	draw=black},
				signal/.style=		{arrows={-stealth},draw=black}]
	\def\nodedist{35pt}
	\def\layerdist{60pt}
    
	\foreach \y in {1,...,3}
		\node[inputnode] (I\y) at (0,-\y*\nodedist) {$x_\y$};  
	\foreach \y in {1,...,4}
		\node[hiddennode] (H1\y) at ($(\layerdist,-\y*\nodedist) +(0, 0.5*\nodedist)$) {};
	\foreach \y in {1,...,4}
		\node[hiddennode] (H2\y) at ($(2*\layerdist,-\y*\nodedist) +(0, 0.5*\nodedist)$) {};
	\foreach \y in {1,...,2}
		\node[outputnode] (O\y) at ($(H21) + (\layerdist, -\y*\nodedist)$) {$y_\y$};

	\foreach \dest in {1,...,4}
		\foreach \source in {1,...,3}
			\draw[signal] (I\source) -- (H1\dest);
	\foreach \dest in {1,...,4}
		\foreach \source in {1,...,4}
			\draw[signal] (H1\source) -- (H2\dest);
	\foreach \dest in {1,...,2}
		\foreach \source in {1,...,4}
			\draw[signal] (H2\source) edge (O\dest);
\end{tikzpicture}
\caption{\textbf{Fully connected neural network with two hidden layers}.}
\label{fig:simple_network}
\end{figure}
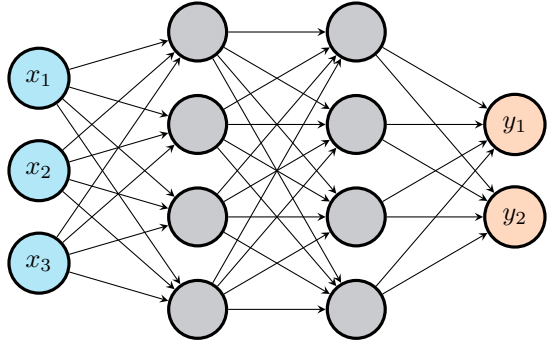

Deep reinforcement learning (or DRL) is the result of applying \blue{reinforcement learning} with deep neural networks to output either value functions (value-based RL methods), or action distributions given input states (policy-based RL methods)\footnote{An alternative presented above is to use tables to store the values for every state or state-action pair, but such a strategy generally does not scale with the size of state-action spaces, and is thus limited to discrete spaces.}. 
A neural network is a collection of artificial neurons, \textit{i.e.} connected computational units with universal approximation capabilities \cite{hornik1989,siegelmann1995}, that can be trained to arbitrarily well approximate the mapping function between input and output spaces. Each connection provides the output of a neuron as an input to another neuron. Each neuron performs a weighted sum of its inputs, to assign significance to the inputs with regard to the task the algorithm is trying to learn. It then adds a bias to better represent the part of the output that is actually independent of the input. Finally, it feeds a non-linear activation function that determines whether and to what extent the computed value should affect the ultimate outcome. As sketched in figure \ref{fig:simple_network}, a fully connected network is generally organized into layers, with the neurons of one layer being connected solely to those of the immediately preceding and following layers. The layer that receives the external data is the input layer, the layer that produces the outcome is the output layer, and in between them are zero or more hidden layers.

The learning process in neural networks consists in adjusting all the biases and weights of the network in order to reduce the value of a well-chosen loss function that represents the quality of the network prediction. This update is usually performed by a stochastic gradient method, in which the gradients of the loss function with respect to the weights and biases (\textit{i.e.} the parameterization $\theta$ of the neural network) are obtained using a back-propagation algorithm. The abundant literature available on this topic (see \cite{goodfellow2017} and the references therein) points out that a relevant network architecture (\eg type of network, depth, width of each layer), finely tuned hyper parameters (\textit{i.e.} parameters whose value cannot be estimated from data, \eg, optimizer, learning rate, batch size) and a sufficiently large amount of data to learn from are key ingredients for a successful network training.

\subsection{\blue{Deep reinforcement learning} algorithms}

This section briefly reviews some of the most popular DRL methods encountered in the field of DRL for fluid mechanics. Only their main features are reviewed here, the reader interested in further details (or in the numerous custom variations introduced in the RL literature) is referred to the various references given in this section. 

\subsubsection{Deep Q-networks (DQN)}
\label{section:dqn}

In Q-learning methods, obtaining a converged Q-table for large state and actions spaces can be particularly expensive in terms of environment interactions. To overcome this issue, the map $\mathcal{S}^{+} \times \mathcal{A} \longrightarrow \mathbb{R}$ is represented by a neural network, called deep Q-network \cite{dqn}, tasked with providing an estimate of the Q-value for each possible action given an input state. To do so, the Q-network is trained on state-action-reward transitions obtained by interacting with the environment. The loss used for the training is classically obtained from the Bellman equation (\ref{eq:bellman}):

\begin{equation}
\label{eq:dqn_loss}
	L(\theta) = \underset{(s,a,r,s') \sim \mathcal{D}}{\mathbb{E}} \left[ \frac{1}{2} \left( \left[ r(s,a) + \gamma \max_{a'} Q_\theta (s',a') \right] - \,Q_\theta(s,a) \right)^2 \right],
\end{equation}

where $Q_\theta(s,a)$ is the Q-value \emph{estimate} provided by the DQN for action $s$ and state $a$ under network parameterization $\theta$, and $\mathcal{D}$ represents the set of transitions collected from the environment. The quantity $r(s,a) + \gamma \max_{a'} Q_\theta (s',a')$, denoted \emph{target}, also appears in the Bellman equation (\ref{eq:bellman}), as the estimate $Q_\theta(s, a)$ is equal to the target (and $L(\theta)$ is thus zero) when the optimal set of parameters $\theta^*$ is reached.

In order to balance the trade-off between exploration and exploitation, the DQN algorithm classically implements a stochastic exploration strategy called $\epsilon$-greedy: before each action, a random parameter $p$ is drawn in $\left[0, 1\right]$ and compared to a user-defined value $\epsilon \in \left[0, 1\right]$, and the action prescribed by $\max_a Q_\theta(s, a)$ is taken only if $p>\epsilon$ (otherwise a random action is taken). The value of $\epsilon$ usually decreases during the learning process, thereby progressively reducing exploration in favour of exploitation. Nevertheless, the performance of vanilla DQN remains limited. This has led to multiple developments aimed at stabilizing learning and at improving performance, a handful of which have become standard practice \textit{e.g.}, replay \cite{expreplay}, target networks \cite{dqn}, or double deep Q-networks \cite{ddqn}. For the sake of clarity, we shall not go into the specifics of these evolutions, for which the interested reader can instead refer to \cite{sutton2018} and the references therein.

\subsubsection{Vanilla deep policy gradient}
\label{section:vpg}

In policy methods, a stochastic gradient algorithm is used to perform network updates from the policy loss:

\begin{equation}
\label{eq:pg_loss}
	L(\theta) = \underset{\tau \sim \pi_\theta}{\mathbb{E}} \left[ \sum_{t=0}^T \log \left( \pi_\theta (a_t \vert s_t) \right) R(\tau) \right].
\end{equation}

whose gradient is equal to the policy gradient (\ref{eq:policy_gradient}). The latter is computed with the back-propagation algorithm with respect to each weight and bias by the chain rule, one layer at the time from the output to the input layer. Such a method is also known as Monte Carlo policy gradient, as the loss (\ref{eq:pg_loss}) takes the form of an expected value, that can be numerically calculated using an empirical average over a set of full trajectories.
However, if some low-quality actions are taken along the trajectory, their negative impact will be averaged by the high-quality actions and will remain undetected. This problem can be overcome using actor-critic methods, in which a Q-function evaluation is used in conjunction with a policy optimization.

\subsubsection{Advantage actor-critic (A2C)}
\label{section:a2c}

Different strategies are available to alleviate the high variance of training the agent from (\ref{eq:pg_loss}), for which it has become customary to replace the discounted cumulative reward by the \emph{advantage function}:

\begin{equation*}
	A(s,a) = Q(s,a) - V(s),
\end{equation*}

that represents the improvement in the expected cumulative reward when taking action $a$ in state $s$, compared to the average of all possible actions taken in state $s$. As a result, the loss function reads:

\begin{equation*}
	L(\theta) = \underset{\tau \sim \pi_\theta}{\mathbb{E}} \left[ \sum_{t=0}^T \log \left( \pi_\theta (a_t \vert s_t) \right) A^{\pi_\theta} (s_t, a_t) \right].
\end{equation*}

In practice, the classical policy network (called \emph{actor}) is used concurrently with a second network (called \emph{critic}), that learns to predict the state-value function $V(s)$. The advantage function is then approximated as 

\begin{equation*}
	A(s_t,a_t) \sim r(s_t,a_t) + \gamma V(s_{t+1}) - V(s_t).
\end{equation*}

to avoid having a third network learn to predict the state-action value $Q(s,a)$.
In contrast to the Monte Carlo-style update of vanilla policy gradient methods, the actor-critic algorithm allows training the policy network in a temporal-difference manner, meaning that updates can be performed several times during an episode, thanks to the critic state-value estimate \cite{a2c}.

\subsubsection{Trust-region and proximal policy optimization (TRPO and PPO)}
\label{section:ppo}

The performance of policy gradient methods is hurt by the high sensitivity to the learning rate, \textit{i.e.}, the size of the step to be taken in the gradient direction. Indeed, small learning rates are detrimental to learning, but large learning rates can lead to a performance collapse if the agent falls off the cliff and restarts from a poorly performing state with a locally bad policy (an issue magnified by the fact that the learning rate cannot be tuned locally).
Trust region policy optimization (TRPO \cite{trpo})) ensures continuous improvement by leveraging second-order natural gradient optimization to update the policy parameters within a trust-region of fixed maximum Kullback-Leibler divergence between previous and current policies. Proximal policy optimization (PPO \cite{ppo}) uses a simpler yet effective heuristic to similarly avoid destructive updates. Namely, it relies on the clipped surrogate loss:

\begin{equation*}\label{eq:ppoloss}
	L(\theta) = \underset{(s,a) \sim \pi_{\theta_\text{old}}}{\mathbb{E}} \left[ \min \left( \frac{\pi_\theta (a \vert s)}{\pi_{\theta_{old}} (a \vert s)} , g \left( \epsilon, A^{\pi_{\theta_\text{old}}} (s,a) \right) \right) A^{\pi_{\theta_\text{old}}} (s,a) \right],
\end{equation*}

where

\begin{equation*}
	g(\epsilon, A) = 
	\begin{cases}
		(1+\epsilon) A 	& \text{if } A \geq 0,\\
		(1-\epsilon) A 	& \text{if } A < 0,
	\end{cases}
\end{equation*}

and $\epsilon$ is the clipping range, a small, user-defined parameter defining how far away the new policy is allowed to go from the old one. The general picture is that a positive (resp. negative) advantage increases (resp. decreases) the probability of taking action $a$ in state $s$, but always by a proportion smaller than $\epsilon$, otherwise the min kicks in (\ref{eq:ppoloss}) and its argument hits a ceiling of $1 +\epsilon$ (resp. a floor of $1 - \epsilon$). This prevents stepping too far away from the current policy, and ensures that the new policy will behave similarly.

Due to its improved learning stability and its relatively robust behaviour with respect to hyper-parameters, the PPO algorithm has received considerable attention in the DRL community. As shown in table \ref{table:method_usage}, it is by far the most common DRL algorithm exploited in the context of DRL-based control for fluid dynamics.

\subsubsection{Deep deterministic policy gradient (DDPG)}
\label{section:ddpg}

Deep deterministic policy gradient (DDPG) can be thought as a \blue{Deep Q-network} algorithm for continuous actions spaces, that combines the learning of a Q-network $Q_\theta (s,a)$ (as in the DQN algorithm) and a deterministic policy network $\mu_\phi (s)$. As in DQN, the replay buffer and target network tricks are used, the latter being a key ingredient of the method. Looking back at the DQN loss (\ref{eq:dqn_loss}), it is obvious that the $\max_{a'} Q_\theta (s',a')$ term does not make sense in the context of a continuous action space. In DDPG, the latter is thus approximated using the target network, yielding the modified loss:

\begin{equation}
\label{eq:ddpg_critic_loss}
	L(\theta) =  \underset{(s,a,r,s') \sim \mathcal{D}}{\mathbb{E}} \left[ \frac{1}{2} \left( \left[ r(s,a) + \gamma \, Q_{\theta_\text{targ}} (s',\mu_{\phi_\text{targ}} (s')) \right] - \,Q_\theta(s,a) \right)^2 \right].
\end{equation}

Hence, the policy $\mu_\phi(s)$ is expected to produce actions corresponding to a maximum value predicted by the Q-network, and therefore its loss is obtained straightforwardly as:

\begin{equation}
\label{eq:ddpg_actor_loss}
	L(\phi) =  \underset{s \sim \mathcal{D}}{\mathbb{E}} \left[ Q_\theta \left( s, \mu_\phi(s) \right) \right].
\end{equation}

Finally, a gaussian noise is usually applied to the predicted actions in order to achieve an efficient balance between exploration and exploitation.

\subsubsection{\blue{Twin-delayed deep deterministic policy gradient} (TD3)}
\label{section:td3}

The \blue{twin-delayed deep deterministic policy gradient} (TD3) algorithm is a refinement of the DDPG method that improves its learning stability and robustness against hyper-parameters \cite{td3}. For the sake of brevity, only the differences between the two methods are pointed out here, namely:

\begin{enumerate}
	\item the use of a second Q-network to avoid the common problem of overestimation of the Q-value, as is done in DDQN \cite{ddqn};
	\item additional delays in the policy and target network updates;
	\item additional noises in the target actions.
\end{enumerate}

Compared to standard DDPG, these three modifications largely improve the stability and performance of the method. Yet, as shown in table \ref{table:method_usage}, these two methods have received little attention in the field of DRL-based control for fluid dynamics.

\subsubsection{Soft actor-critic (SAC)}
\label{section:sac}

The soft actor-critic (SAC) algorithm shares similar traits with the TD3 algorithm, but presents two major differences with the latter:

\begin{enumerate}
	\item it exploits a stochastic policy and not a deterministic one;
	\item it maximizes a trade-off between the expected return and the policy entropy, thus efficiently balancing exploration and exploitation.
\end{enumerate}

In the present review, a single article exploits this technique, as shown in table \ref{table:method_usage}.

\subsection{Single-step \blue{deep reinforcement learning}}
\label{section:pbo}

In several contributions assessed in this review, the optimal policy
to be learnt by the neural network is state-independent, as is notably the case in optimization and open-loop control problems. We group here under the "single-step DRL" label the class of algorithms dedicated to this class of problems under the premise that it may be enough for the neural network to get only one attempt per episode at finding the optimal.
In essence, the proposed methods inherit from deep policy gradient algorithms in the sense that relevant probability density function parameters are obtained from neural networks trained using a policy gradient-like loss. Yet, they also fall heir of evolutionary strategies (ES), as their successive steps follow a generation/individual nomenclature, exploiting information from previous generations in order to update the parameters of a probability density function. 
The seminal PPO-1 algorithm proceeds from the standard PPO algorithm (section \ref{section:ppo}) and samples actions isotropically from scalar covariance matrices \cite{ghraieb2020, hachem2020, viquerat2021}. The follow-up policy-based optimization (PBO) algorithm relies on a variant of the vanilla policy gradient method and delivers several major improvements by adopting key heuristics from the covariance matrix adaptation evolution strategy (CMA-ES \cite{cmaes}), including the use of a valid, full covariance matrix generated from neural network outputs \cite{pbo}. \blue{It is noted here that adjective "deep" is used more as an extension of the original method name than as a real description of the networks depth. Indeed, due to the use of a fixed input state, these techniques exploit extremely small networks, with only a few tens of parameters in total.}

\section{Open challenges}
\label{section:challenges}

Before delving into the specifics of the compiled papers, it is important to define a consistent list of challenges to serve as a common thread to measure the progress of DRL in the context of fluid mechanics applications (and also to examine the willingness of the community to take on these challenges). For those challenges left mostly unanswered, section~\ref{section:transverse} proposes a series of possible mitigation strategies that have received consideration in the literature, albeit in a different context. The retained challenges are computational cost (more generally, sampling-efficiency), turbulence, robustness, partial observability, delays, and any combination of them. Nonetheless, there are several other challenges that should be considered on the second level to help bridge the gap between DRL capabilities and the requirements of practical deployment, for instance multi-agent DRL (leveraging experience from multiple agents learning concurrently) or multi-objective reward (training an agent in reasoning about several weighted objectives), see, \eg \cite{dulac2019, dulac2021, garau2021} for comprehensive domain-agnostic surveys.

\textit{$\circ$ Computational cost/sampling efficiency:} the environment of computational fluid dynamics (CFD) problems is resource expensive, as it routinely involves numerical simulations with tens or hundreds of millions of degrees of freedom (unless an appropriate low-dimensional reduction is achieved, which in itself often proves very challenging). This is all the more problematic since classical RL methods have low sample efficiency, \textit{i.e.} many trials are required for the agent to learn a purposive behavior.

\textit{$\circ$ Stochasticity (turbulence):} most natural and engineered flows are turbulent and carry energy distributed over a wide range of scales with varying degrees of spatial and temporal coherence. Their dynamics therefore inherently includes some degree of stochasticity, which might lead to high variance gradient estimates that hamper learning.

\textit{$\circ$ Robustness:} optimizing robust policies is a key issue for fluid flow applications, with multiple sources of uncertainty relating to the occurrence of irregular transient dynamics and the high sensitivity to initial conditions and system parameters variations (all non-normal amplification mechanisms associated with the asymmetry of the Navier--Stokes convection operator), not to mention the difficulty to consistently ascertain the accuracy of the computed numerical solutions.

\textit{$\circ$ Partial observability:} the traditional states space of fluid flow problems are easily prohibitively large for policy learning. The agent must therefore operate under partially observable environments, in which case the performance becomes highly dependent on the quality and relevance of the data available for observation. This issue is strongly related to data-driven model reduction techniques for large scale dynamical systems, which usually require using measures of observability as an information quality metric.

 \textit{$\circ$ Pre-action delays:} in numerical environments, states are collected in the environment, provided to the agent, and actions are returned  instantaneously, which amounts to artificially interrupting the lapse of time every time the agent must draw new actions. In real-world environments, a certain delay is inevitable due to data processing, data transition, and physical constraints of sensors and actuators, during which the environment keeps evolving, meaning that the agent actually takes actions based on out-dated states.

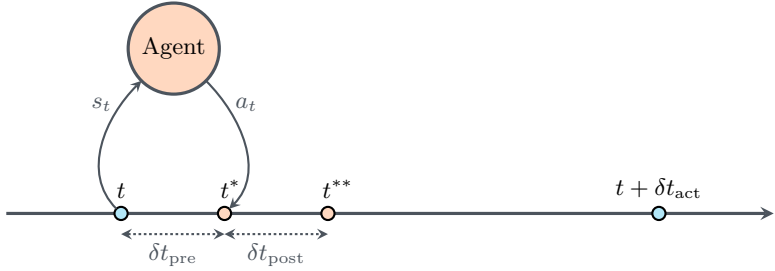
\begin{figure}
\centering
\def\myscale{0.55}
\def\sx{10}
\begin{tikzpicture}[	scale=\myscale,
				fontsize/.style=		{font=\footnotesize},
				tick/.style=		{circle, inner sep=0pt, outer sep=0pt, text width=5pt, thick, fill=blue4, draw=black},
				timeline/.style= 		{-stealth,gray1,very thick},
				arrow/.style=		{-stealth,gray1,thick},
				box/.style={		circle, draw=gray1, very thick, 
								text width=1cm,minimum height=1cm,text centered,
								inner sep=2pt, outer sep=0pt, fontsize}]
				
	\node (start) at (-3,0) {};
	\node (end) at (16,0) {};
	\draw[timeline] (start) -- (end);
	
	\node[tick] (tn) 		at  (0,0) {};
	\node[fontsize, above=of tn, yshift=-1cm] (tnt)	{$t$};
	\node[tick] (tnp) 	at (13,0) {};
	\node[fontsize, above=of tnp, yshift=-1cm] (tnpt)	{$t+\delta t_\text{act}$};
	\node[tick,fill=orange4] (ts) at  (2.5,0) {};
	\node[fontsize, above=of ts, xshift=0.08cm,yshift=-1cm] (tst)	{$t^*$};
	\node[tick,fill=orange4] (tss) at  (5,0) {};
	\node[fontsize, above=of tss, xshift=0.13cm,yshift=-1cm] (tst)	{$t^{**}$};
	
	\node[box,fill=orange4,above=of tn,xshift=0.7cm,yshift=0.5cm] (agent) {Agent};
	
	\draw[arrow] (tn.north west) to [out=135,in=225] node[fontsize,pos=0.8,left] (st) {$s_t$} (agent.south west);
	\draw[arrow] (agent.south east) to [out=-45,in=45] node[fontsize,pos=0.2,right] (at) {$a_t$} (ts.north east);
	\draw[arrow,stealth-stealth,dash pattern=on 1pt] ($(tn) + (0.0cm,-0.5cm)$) to [out=0,in=180] node[fontsize,pos=0.5,below] {$\delta t_\text{pre}$} ($(ts) + (-0.0cm,-0.5cm)$);
	\draw[arrow,stealth-stealth,dash pattern=on 1pt] ($(ts) + (0.0cm,-0.5cm)$) to [out=0,in=180] node[fontsize,pos=0.5,below] {$\delta t_\text{post}$} ($(tss) + (-0.0cm,-0.5cm)$);
	
\end{tikzpicture}
\caption{\textbf{Different type of delays encountered in \blue{deep reinforcement learning} environments.} \emph{Pre-action delay} corresponds to the time delay existing between the moment of state collection and the moment when action is applied to the environment (this type of delay does not exist in the case of numerical environments). \emph{Post-action delay} is defined as the time delay existing between the moment the action is applied to the environment, and the moment where it becomes fully effective in the dynamics of the system.}
\label{fig:delays}
\end{figure}

\textit{$\circ$ Post-action delays:} an environment has an intrinsic response time that depends on the interplay between transient amplification of the action-induced initial energy and non-linear saturation (the former all the more important in fluid mechanics where non-normal systems are common occurrence). This entails \emph{post-action delays}, defined as the time interval between the moment an action is applied, and the moment it efficiently reaches the current state, that can undermine the accuracy of the reward estimation and even prevent learning if they exceed the Lyapunov time (the characteristic time scale on which a dynamical system is chaotic). The two types of delays defined above are illustrated in figure \ref{fig:delays}, the general picture being that post-action delays affect both numerical or experimental environment, while pre-action delays affect only experimental environments (unless they are purposely included in a numerical model).

\begin{table}
\caption{\textbf{Classification of the reviewed papers by domain of application.} The most represented domain of application is drag reduction, with no less than \green{18} papers in total.}
\label{table:papers}
\centering
\medskip
\begin{tabular}{ccc}																																	
\toprule
\textbf{Category} 					& \textbf{Domain}		& \textbf{Reference}																												\\\cmidrule(lr){1-3}
\multirow{6}{*}[-1.3em]{Numerical} 		& Drag reduction		& \cite{koizumi2018, rabault2019, rabaultkuhnle2019, tokarev2020, xu2020, tang2020, elhawary2020, holm2020, paris2020, ghraieb2020, qinwang2021, ren2021, li2021, castellanos2022, yzwang2022, pino2022, mei2022, mao2022}	\\\cmidrule(lr){2-3}
								& Heat transfer			& \cite{beintema2020, hachem2020}																									\\\cmidrule(lr){2-3}
								& Microfluidics			& \cite{lee2019, dressler2018}																										\\\cmidrule(lr){2-3}
								& Swimming			& \cite{novati2017, verma2018, yan2020, zhu2021, yan2021, zhu2022}																						\\\cmidrule(lr){2-3}
								& Shape optimization	& \cite{yan2019, li2020, viquerat2021, qin2021}																							\\\cmidrule(lr){2-3}
								& Other				& \cite{ma2018, belus2019, wei2019, xie2021, novati2021, zheng2021, qwang2022, kim2022}																						\\\cmidrule(lr){1-3}
\multirow{3}{*}[-0.4em]{Experimental}	& Drag reduction		& \cite{fan2020}																												\\\cmidrule(lr){2-3}
								& Flow separation 		& \cite{shimomura2020}																											\\\cmidrule(lr){2-3}
								& Microfluidics 			& \cite{dressler2018}																												\\\cmidrule(lr){1-3}
Review							& -					& \cite{garnier2021, rabault2020}																									\\
\bottomrule
\end{tabular}
\end{table}

\section{\blue{Deep reinforcement learning} for computational fluid dynamics}
\label{section:revcfd}

Of the \green{43} papers compiled in the present review, \green{40} consider applying DRL to computational fluid dynamic (CFD) systems. Those are classified and presented here in one of the categories listed in table \ref{table:papers}, to put similar papers in perspective with respect to one another and to point out their specificities.

\subsection{Drag reduction}
\label{section:drag_reduction}

\begin{figure}
\centering
\def\xaxis{4}
\def\rad{1}
\def\sc{0.8}
\begin{subfigure}[t]{.2\textwidth}
	\centering
	\begin{tikzpicture}[	scale=\sc,
					center/.style={draw=none, inner sep=0pt, outer sep=-1pt},
					line/.style={draw=black},
					jet/.style={draw=bluegray1, thick, -{Stealth[length=1mm,width=1mm]}},
					jetline/.style={draw=bluegray1, thick},]

		\draw[black, very thick] (0,0) rectangle (\xaxis,\xaxis);
	
		\filldraw[fill=blue4,draw=blue1,thick] (0.5*\xaxis,0.5*\xaxis) circle (\rad);
		
		\foreach \y in{0.1,0.2,0.3,0.4,0.5,0.6,0.7,0.8,0.9}
			\draw[-stealth, orange1] (0,\y*\xaxis) -- (0.5,\y*\xaxis);
		
		\node[center] (c) at (0.5*\xaxis,0.5*\xaxis) {};

		\draw[line] (c) -- ($(c) + (0.1736*\rad,0.9848*\rad)$);
		\draw[line] (c) -- ($(c) + (-0.1736*\rad,0.9848*\rad)$);
		\draw[jetline] (c) ++(79:1)  arc(79:101:1) node[pos=0.5, anchor=south, yshift=0.2cm] {\scriptsize $q_1$};
		
		\draw[jet] ($(c) + (0.087*\rad,0.996*\rad)$) -- ($(c) + (0.087*1.2*\rad,0.996*1.2*\rad)$);
		\draw[jet] ($(c) + (0.0*\rad,1*\rad)$) -- ($(c) + (0*1.3*\rad,1*1.3*\rad)$);
		\draw[jet] ($(c) + (-0.087*\rad,0.996*\rad)$) -- ($(c) + (-0.087*1.2*\rad,0.996*1.2*\rad)$);
		
		\draw[line] (c) -- ($(c) + (0.1736*\rad,-0.9848*\rad)$);
		\draw[line] (c) -- ($(c) + (-0.1736*\rad,-0.9848*\rad)$);
		\draw[jetline] (c) ++(-79:1)  arc(-79:-101:1) node[pos=0.5, anchor=north, yshift=-0.2cm] {\scriptsize $q_2$};
		
		\draw[jet] ($(c) + (0.087*\rad,-0.996*\rad)$) -- ($(c) + (0.087*1.2*\rad,-0.996*1.2*\rad)$);
		\draw[jet] ($(c) + (0.0*\rad,-1*\rad)$) -- ($(c) + (0*1.3*\rad,-1*1.3*\rad)$);
		\draw[jet] ($(c) + (-0.087*\rad,-0.996*\rad)$) -- ($(c) + (-0.087*1.2*\rad,-0.996*1.2*\rad)$);

	\end{tikzpicture}
	\caption{Lateral zero-mass-flow-rate jets.}
	\label{fig:drag_reduction_jets}
\end{subfigure} \qquad
\begin{subfigure}[t]{.2\textwidth}
	\centering
	\begin{tikzpicture}[	scale=\sc,
					center/.style={draw=none, inner sep=0pt,outer sep=0pt},
					rot/.style={draw=bluegray1, thick, -{Stealth[length=1mm,width=1mm]}}]

		\draw[black, very thick] (0,0) rectangle (\xaxis,\xaxis);
	
		\filldraw[fill=blue4,draw=blue1,thick] (0.5*\xaxis,0.5*\xaxis) circle (\rad);
		
		\foreach \y in{0.1,0.2,0.3,0.4,0.5,0.6,0.7,0.8,0.9}
			\draw[-stealth, orange1] (0,\y*\xaxis) -- (0.5,\y*\xaxis);
		
		\node[center] (c) at (0.5*\xaxis,0.5*\xaxis) {};
		
		\draw[rot] (c) ++(30:1.2)  arc(30:-30:1.2) node[pos=0.5,anchor=west] {\scriptsize $\omega$};
    
	\end{tikzpicture}
	\caption{Main cylinder rotating.}
	\label{fig:drag_reduction_rotating}
\end{subfigure} \qquad
\begin{subfigure}[t]{.2\textwidth}
	\centering
	\begin{tikzpicture}[	scale=\sc,
					center/.style={draw=none, inner sep=0pt,outer sep=0pt},
					rot/.style={draw=bluegray1, thick, -{Stealth[length=1mm,width=1mm]}}]

		\draw[black, very thick] (0,0) rectangle (\xaxis,\xaxis);
	
		\filldraw[fill=blue4,draw=blue1,thick] (0.5*\xaxis,0.5*\xaxis) circle (\rad);
		
		\foreach \y in{0.1,0.2,0.3,0.4,0.5,0.6,0.7,0.8,0.9}
			\draw[-stealth, orange1] (0,\y*\xaxis) -- (0.5,\y*\xaxis);
		
		\node[center] (c) at (0.5*\xaxis,0.5*\xaxis) {};
		\node[center] (c1) at (0.5*\xaxis+1.25,0.5*\xaxis+1.0) {};
		\node[center] (c2) at (0.5*\xaxis+1.25,0.5*\xaxis-1.0) {};
		
		\filldraw[fill=blue4,draw=blue1,thick] (c1) circle (0.2*\rad);
		\filldraw[fill=blue4,draw=blue1,thick] (c2) circle (0.2*\rad);
		
		\draw[rot] (c1) ++(130:0.3*\rad)  arc(130:50:0.3*\rad) node[pos=0.5,anchor=south] {\scriptsize $\omega_1$};
		\draw[rot] (c2) ++(-130:0.3*\rad)  arc(-130:-50:0.3*\rad) node[pos=0.5,anchor=north] {\scriptsize $\omega_2$};
    
	\end{tikzpicture}
	\caption{Downstream rotating control cylinders.}
	\label{fig:drag_reduction_downstream_rotating}
\end{subfigure} \qquad
\begin{subfigure}[t]{.2\textwidth}
	\centering
	\begin{tikzpicture}[	scale=\sc,
					center/.style={draw=none, inner sep=0pt,outer sep=-0.5pt},
					line/.style={draw=black},
					lined/.style={line, dash pattern=on 1pt},
					jetline/.style={draw=bluegray1, thick},
					rot1/.style={draw=bluegray1, thick, {Stealth[length=1mm,width=1mm]}-},
					rot2/.style={rot1, lined, thin}]

		\draw[black, very thick] (0,0) rectangle (\xaxis,\xaxis);
	
		\filldraw[fill=blue4,draw=blue1,thick] (0.5*\xaxis,0.5*\xaxis) circle (\rad);
		
		\foreach \y in{0.1,0.2,0.3,0.4,0.5,0.6,0.7,0.8,0.9}
			\draw[-stealth, orange1] (0,\y*\xaxis) -- (0.5,\y*\xaxis);
		
		\node[center] (c) at (0.5*\xaxis,0.5*\xaxis) {};
		
		\draw[line] (c) -- ($(c) + (0.5*\rad,0.866*\rad)$); 
		\draw[line] (c) -- ($(c) + (0.259*\rad,0.965*\rad)$); 
		\draw[lined] (c) -- ($(c) + (0.382*\rad,0.924*\rad)$); 
		\draw[jetline] (c) ++(59:\rad)  arc(59:76:\rad);
		\draw[rot1] (c) ++(60:1.1*\rad)  arc(60:75:1.1*\rad) node[pos=0.5,anchor=south] {\scriptsize $q_1$};
		
		\draw[line] (c) -- ($(c) + (0.5*\rad,-0.866*\rad)$); 
		\draw[line] (c) -- ($(c) + (0.259*\rad,-0.965*\rad)$); 
		\draw[lined] (c) -- ($(c) + (0.382*\rad,-0.924*\rad)$); 
		\draw[jetline] (c) ++(-59:\rad)  arc(-59:-76:\rad);
		\draw[rot1] (c) ++(-60:1.1*\rad)  arc(-60:-75:1.1*\rad) node[pos=0.5,anchor=north] {\scriptsize $q_2$};
    
	\end{tikzpicture}
	\caption{Symmetric plasma actuators.}
	\label{fig:drag_reduction_plasma}
\end{subfigure}
\caption{\textbf{Different drag reduction methods represented in the \blue{deep reinforcement learning} literature, in the context of moderate Reynolds flows around a 2D circular cylinder.} (\ref{fig:drag_reduction_jets}) Zero-mass-flow-rate jets are used to blow or suck fluid on the lateral sides of the obstacle. There can be two or four, possibly tilted. (\ref{fig:drag_reduction_rotating}) An angular velocity is applied to the obstacle, in order to alter the downstream flow and reduce drag. (\ref{fig:drag_reduction_downstream_rotating}) Two small control cylinders, placed downstream of the obstacle, are given angular velocities in order to stabilize the shedding of the main cylinder. (\ref{fig:drag_reduction_plasma}) Two symmetric plasma actuators are controlled to alter the fluid flow near the flow-separation point, thus reducing the overall drag on the obstacle.}
\label{fig:drag_reduction_methods}
\end{figure}
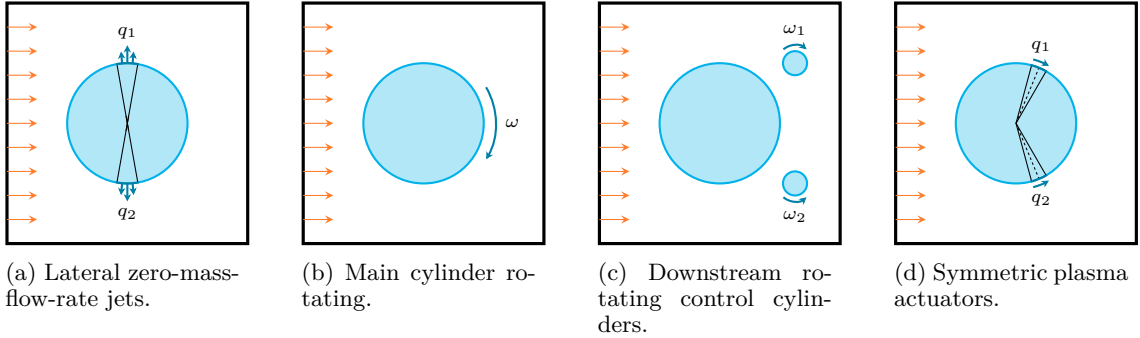

Drag reduction is by far the most represented application domain in the literature, with 16 different papers implementing various control strategies using zero-mass-flow-rate jets \cite{rabault2019, rabaultkuhnle2019, koizumi2018, tang2020, ren2021, paris2020, qinwang2021, li2021, castellanos2022, yzwang2022, pino2022, mei2022, mao2022}, rotating cylinders \cite{tokarev2020, xu2020, holm2020, ghraieb2020}, plasma actuators \cite{elhawary2020}, or passive devices \cite{ghraieb2020}, as illustrated in figure \ref{fig:drag_reduction_methods}. Almost all studies focus on prototypal, two-dimensional (2-D) incompressible flows past span-wise infinite cylinders (generally different sections of a single cylinder) subjected to a uniform and/or parabolic velocity profile, at the exception of \cite{mei2022}, which considers cylinders with variable cross-sections, and \cite{yzwang2022}, which considers the case of a NACA airfoil, again in a parabolic flow. As seen from the comparison between studies provided in table \ref{table:drag_reduction_summary}, most DRL algorithms belong to the actor-critic category, with a clear preference for ready-to-use PPO implementations (see section \ref{section:ppo}), either from Tensorforce \cite{tensorforce}, OpenAI baselines \cite{baselines}, or Stable Baselines \cite{stable-baselines}. Regarding the CFD solvers, FeniCS \cite{fenics} is well represented, mostly because the open-source diffusion of the seminal work from Rabault \textit{et al.\@} \cite{rabault2019} has been heavily re-used in follow-up works \cite{rabaultkuhnle2019, xu2020, tang2020, elhawary2020, holm2020, qinwang2021, castellanos2022, pino2022}. Regarding the numerical implementation, since performing a relevant network update requires evaluating a sufficient number of actions drawn from the current policy (which in turn requires computing the same amount of rewards from resource-expensive numerical simulations), most studies have the agent acquire experience at a faster pace by interacting with multiple environments simultaneously. This has become a customary procedure after the methodological paper by Rabault and Kuhnle \cite{rabaultkuhnle2019} on the accelerated gathering of state-action-reward transitions, which highlighted an almost perfect speedup up to 20 parallel environments, and a decent performance improvement up to 60.

 \begin{sidewaystable}
\begin{center}
\caption{\textbf{Summary of the main features of \blue{deep reinforcement learning} drag reduction applications.} Only explicitly stated informations were retained from the contributions. Missing informations are noted by a question mark "?", while non-applicable data are noted with a double-dash "--". In the case where multiple studies were conducted in the considered paper with different parameter values, the most significant one was retained. $n_\text{probes}$ corresponds to the number of sensors placed in the domain for observation collection, while $n_\text{act}$ represents the action dimensionality. The information in the actor column refers to fully-connected network architecture used (in most contributions, the critic architecture is missing). Finally, the information in the last two columns pertain respectively to the ratio $\delta t_\text{act}/\delta t$ of the action to the simulation time-steps, and to the ratio $\delta t_\text{phy}/\delta t_\text{act}$ of the physical time scale (here equal to the vortex shedding period) to the action time-step. The acronyms for the DRL packages are the following: TFce = TensorForce, OAIb = OpenAI Baselines, StB = Stable Baselines.}
\label{table:drag_reduction_summary}					
\begin{tabularx}{\linewidth}{ccccccccccc}																		
\toprule
\textbf{Control} 	& \textbf{Reference}			& \textbf{Strategy}		& \textbf{$\Re$}	& \textbf{DRL}				& \textbf{CFD}		& \textbf{$n_\text{probes}$}	& \textbf{$n_\text{act}$} 	&	\textbf{Actor}		& \textbf{$\delta t_\text{act}/\delta t$}	& \textbf{$\delta t_\text{phy}/\delta t_\text{act}$} 	\\\cmidrule(lr){1-11}
\multirow{10}{*}[-2.2em]{Active/Closed-loop} 
			& \cite{rabault2019}			&  Jets				& 100		& PPO (\red{TFce})			& FeniCS			& 151 ($v$)				& 1					&	$[512,512]$		& 50							& 12						\\\cmidrule(lr){2-11}
			& \cite{rabaultkuhnle2019} 	& \guillemotright		& 100		& PPO (\red{TFce})			& FeniCS			& 151 ($v$)				& 1					& 	$[512,512]$		& 50							& 12						\\\cmidrule(lr){2-11}
			& \cite{koizumi2018}  		& \guillemotright		& 100		& DDPG (\red{in-house})		& UPACS 			& 1 ($v$)					& 1					&	$[400,300]$		& 100						& 60						\\\cmidrule(lr){2-11}
			& \cite{qinwang2021}		& \guillemotright		& 100		& PPO (\red{TFce})			& FeniCS 			& 63 ($p$)					& 1					&	$[512,512]$		& 50							& 12						\\\cmidrule(lr){2-11} 
			& \cite{paris2020}			& \guillemotright		& 120		& \red{PPO} (in-house)		& FastS 			& 3--16 ($p$)				& 1					&	$[512,512]$		& 50							& 22						\\\cmidrule(lr){2-11} 
			& \cite{tang2020}			& \guillemotright		& 100--400	& PPO (\red{TFce})			& FeniCS			& 236 (?)					& 2					&	$[512,512]$		& 200						& 30						\\\cmidrule(lr){2-11} 
			& \cite{ren2021} 			& \guillemotright		& 1000		& PPO (\red{in-house})		& LBM 			& 151 ($v$)				& 1					&	?				& 300						& --						\\\cmidrule(lr){2-11}
			& \cite{li2021} 				& \guillemotright		& 100		& PPO (\red{TFce})			& Nek5000		& 86 ($v$)					& 1					&	$[512,512]$		& 40 							& 12--16					\\\cmidrule(lr){2-11}
			& \cite{yzwang2022} 			& \guillemotright		& 3000		& PPO (\red{TFce})			& OpFm			& ? ($p,v$)				& 3					&	$[512,512]$		& ? 							& ?						\\\cmidrule(lr){2-11}
			& \cite{pino2022} 			& \guillemotright		& 400		& DDPG (\red{in-house})		& FeniCS			& 5 ($p$)					& 2					&	$[128,128]$		& 100						& 33						\\\cmidrule(lr){2-11}
			& \cite{castellanos2022}		& \guillemotright		& 100		& PPO (\red{TFce})			& FeniCS			& 5--11 ($v$)				& 1					&	$[512,512]$		& 50							& 12						\\\cmidrule(lr){2-11}
			& \cite{mao2022}			& \guillemotright		& \green{400}	& \green{PPO (TFce)}		& \green{FeniCS}	& \green{236 ($v$)}			& \green{3}			&	$\green{[512,512]}$	& \green{100}					& \green{29}				\\\cmidrule(lr){1-11}			
			& \cite{tokarev2020}			& Rotation				& 100		& PPO (\red{OAIb})			& T-Flows 		& 12 ($p$)					& 1					&	$[64,64]$			& 30							& 20						\\\cmidrule(lr){2-11} 
			& \cite{holm2020}			& \guillemotright		& 100--200	& PPO (\red{TFce})			& FeniCS			& 476 ($p$)				& 3					&	$[512,512]$		& 85--350						& 10--20					\\\cmidrule(lr){2-11} 
			& \cite{xu2020}				& \guillemotright		& 240		& PPO (\red{TFce})			& FeniCS			& 99	($v$)					& 2					&	?				& ?							& ?						\\\cmidrule(lr){1-11} 
\multirow{2}{*}[-0.2em]{Active \newline Open-loop}  
			& \cite{elhawary2020}		& Plasma				& 100		& A2C (\red{in-house})		& FeniCS			& 10 ($p$)					& 1					&	$[128,64]$		& ?							& 1						\\\cmidrule(lr){2-11}
			& \cite{ghraieb2020}			& Rotation				& 2200		& PPO-1 (\red{StB})			& Cimlib 			& --						& 2--3				&	$[4,4]$			& --							& --						\\\cmidrule(lr){1-11} 
\multirow{2}{*}[0.6em]{\parbox{2cm}{Passive}}
			& \cite{ghraieb2020}			& Device 				& 100-22000	& PPO-1 (\red{StB})			& Cimlib 			& --						& 2--3				&	$[4,4]$			& --							& --						\\
\bottomrule
\end{tabularx}
\end{center}  
\end{sidewaystable}

\begin{figure}
\centering
\fbox{\includegraphics[width=.6\linewidth]{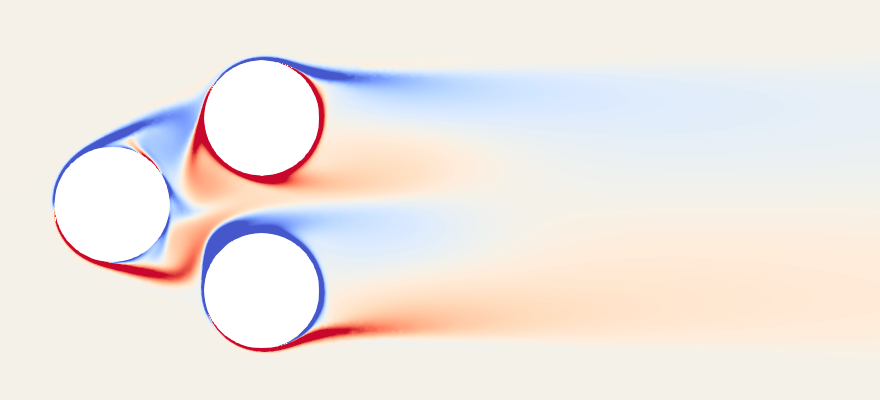}}
\caption{\textbf{Vorticity field of the fluidic pinball case considered in \cite{ghraieb2020}.} The three cylinders are free to rotate at different angular velocities, leading to complex flow features in the near wake region.}
\label{fig:pinball}
\end{figure}

Regarding the flow regimes, almost all contributions assume laminar conditions with Reynolds numbers in a range of one hundred to a few hundred. The only exceptions are \cite{ren2021, yzwang2022}, where weakly turbulent flows at intermediate Reynolds numbers ($\Re = 1000$ and $\Re = 3000$ respectively) are explicitly targeted, and \cite{ghraieb2020}, where moderately large Reynolds number in the range of a few hundreds to a few ten thousands are tackled in the frame of Reynolds averaged Navier--Stokes (RANS) (see figure \ref{fig:pinball} for an illustration pertaining to the fluidic pinball, an equilateral triangle arrangement of rotating cylinders immersed in a turbulent stream). As stressed in \cite{ren2021}, even weakly turbulent conditions make it significantly harder to achieve successful drag reduction, as evidenced by the increased number of episodes needed to learn an efficient policy at higher Reynolds numbers. Moreover, the use of transfer learning from strategies learned at $\Re=100$ to flow control at $\Re=1000$ is shown to be ineffective in this configuration, due to too different flow dynamics. Nonetheless, it is shown possible in \cite{tang2020} to achieve robust flow control over a range of Reynolds numbers by training simultaneously a single agent at four different Reynolds numbers distributed between 100 and 400. After training, the agent succeeds in efficiently reducing drag for Reynolds numbers in the range from 60 to 400, although the performance for each value of $\Re$ is slightly lower than that achieved training an agent specifically at this Reynolds number. \green{Also, the onset of weak turbulence in the shear layer begins to affect the state space observed in the wake and triggers a high level of irregular drag and actuation fluctuations.}
 In \cite{paris2020}, the authors also underline the interest of using non-dimensionalized quantities as input states, which can increase the robustness of control strategies even learned at a single Reynolds number. Yet, as stated above about the work of \cite{ren2021}, this approach is limited to cases with similar flow patterns, and does not carry over to turbulent (not even weakly turbulent) flows. \green{The case at Re=400 has been recently revisited by \cite{mao2022} using an original combination of Markov decision processes with time delays, to manage the time elapsed between the actuation and the flow response, by taking into account previous actuation informations in the agent’s current decision, and autoregressive policy models, to handle the difficult exploration of the agent due to the presence of weak turbulence. In the case of Gaussian policy distribution, the action can be represented as a sum of deterministic parameterized mean and a scaled white noise. It is the white noise that can undermine the exploration behavior, as the mean usually varies smoothly between subsequent states. Autoregressive policy models replace the white noise component with a stationary autoregressive Gaussian process that has stationary standard normal distribution and exhibits temporal coherence between subsequent observations. Such an approach shows promise for turbulent flows as it is reduces the magnitude of drag and lift fluctuations by approximately 90\% while achieving a similar level of drag reduction.}

About the numerical reward, most contributions rely on the design proposed in \cite{rabault2019}:

\begin{equation}
\label{eq:reward_drag_lift}
r_t = - \left< C_D \right> - \beta \left| \left<C_L\right> \right|,
\end{equation}

where the operator $\left< \cdot \right>$ indicates the sliding average over one vortex shedding period $T$ \cite{rabault2019, elhawary2020, paris2020, castellanos2022, pino2022} or over one action time-step $\delta t_\text{act}$ \cite{tokarev2020, ren2021, tang2020, holm2020, yzwang2022}. The parameter $\beta$ varies in a range from 0.2 to 1 in the papers reviewed in this section, and prevents the network to achieve efficient drag reduction by relying on a large induced lift, as it is damageable in many practical applications. Whenever a different reward is used, penalization terms associated with the cost of the control are rarely considered (with reference \cite{ghraieb2020} being an exception), as it has been found customary to explicitly bound the actuation amplitude for the control to remain small compared to the system relevant physical quantities. 
Of particular interest is the recent approach of \cite{qinwang2021} exploiting dynamic mode decomposition to design a reward function based on mode amplitudes, that has led to efficient control strategies, although the proposed approach supposes additional reward tuning compared to (\ref{eq:reward_drag_lift}).

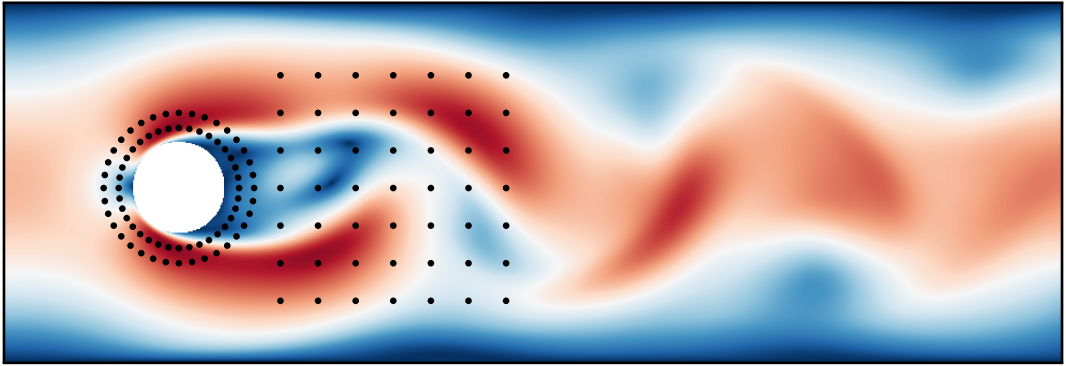
\begin{figure}
\centering
\def\sc{1}
\begin{tikzpicture}[	scale=\sc,
				probe/.style={circle, fill=black, inner sep=0pt, minimum size=2.5pt}]

	\node[anchor=south west,inner sep=0,scale=\sc] (image) at (0,0) {\fbox{\includegraphics[width=0.9\textwidth]{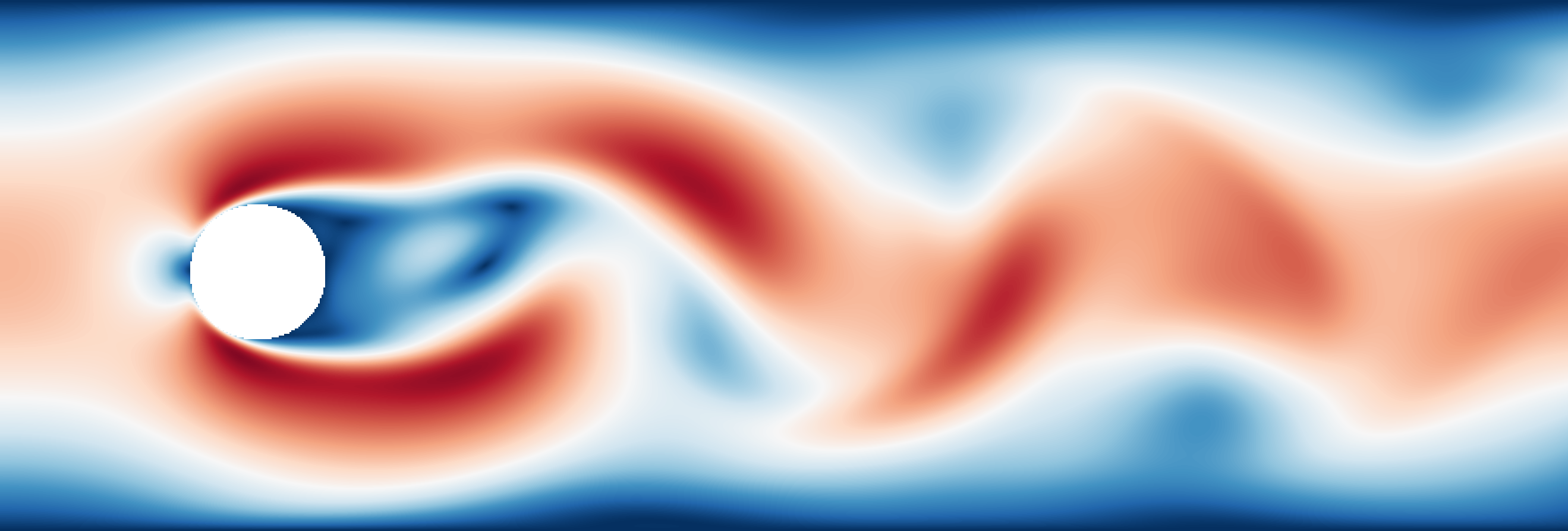}}};

	\foreach \x in {0,...,6}
    		\foreach \y in {0,...,6} 
			\node[probe] at (3.7 + 0.5*\x,0.85 + 0.5*\y) {};
	\foreach \t in {0,10,...,350}
		\node[probe] at ($ (\t:0.8) + (2.35,2.35) $) {};
	\foreach \t in {0,10,...,350}
		\node[probe] at ($ (\t:1.0) + (2.35,2.35) $) {};

\end{tikzpicture}
\caption{\textbf{Typical probe array for observation collection} in the context of drag reduction application on a 2D cylinder at moderate Reynolds numbers. Although the amount and position of probes vary, many contributions position probes in the vicinity of the obstacle and downstream of it. Insights on the impact of this choice regarding the performance of the agent can be found in \cite{rabault2019} and \cite{paris2020}.}
\label{fig:drag_reduction_probes}
\end{figure}

While the considered number of free action parameters $n_\text{act}$ remains limited to \green{4} at most, the number of probes used to collect observations varies considerably from one contribution to another, even for similar setups. The baseline configuration consists of a certain number of velocity or pressure probes, uniformly distributed in the vicinity of the cylinder as well as in its wake region, as illustrated in figure \ref{fig:drag_reduction_probes}. The sensitivity of the learned control strategy to the probes distribution has been briefly studied in \cite{rabault2019}. A subsequent, more complete analysis has been performed in \cite{paris2020}, where the authors evidence a critical impact on the control performance, the information provided by sensors positioned in the near wake being reportedly more relevant for learning than that of sensors positioned further downstream. This can be attributed to the fact that, by observing the flow closely downstream of the cylinder, the agent is able to observe the consequences of its actions right after they were taken, while more distant sensors provide a delayed feedback that can be more difficult to interpret. An extended sensitivity analyses proposed in \cite{li2021} supports these results, showing that more efficient control laws are obtained when data is collected in the areas of high sensitivity. Yet, it most cases (at high $Re$ values, for example), performing a sensitivity analysis of the problem may not be a possible option. To circumvent this issue, the authors in \cite{paris2020} introduce a specific method, called sparse PPO-CMA, in which an optimal set of sensors is automatically selected during the learning process. Another notable approach is that of \cite{koizumi2018}, where only one probe is used downstream of the cylinder, collecting observations at a higher rate than the action time-step $\delta t_\text{act}$, and stacking them into a single observation vector when feeding them to the agent. In all relevant contributions, pressure or velocity are used indifferently as observations. 

As stated previously, the frequency at which the agent is allowed to provide new actions to the environment is defined by an action time-step $\delta t_\text{act}$, that must be larger than the numerical simulation time-step $\delta t$ (for the agent to be able to observe the effects of its actions on the environment), but smaller than the characteristic time scale $\delta t_\text{phy}$ of the physical process to be controlled (for the actions taken to be able to significantly alter the flow dynamics). Considerations about numerical stability and accuracy of the numerical flow solution call for $\delta t \ll \delta t_\text{phy}$, meaning that the user has considerable leeway to adjust the action time step within these two bounds. In the reviewed contributions, the ratio of the action time-step to the physical time scale $(\delta t_\text{act}/\delta t_\text{phy})$ is in a range of a few tens (\ie, a few tens of actions are taken per vortex shedding period). Meanwhile, the ratio of the action time-step to the simulation time-step $(\delta t_\text{act}/\delta t)$ varies considerably with the Reynolds number, as it is set to a few tens at $\Re=100$ \cite{rabault2019, tokarev2020, paris2020}, but increases up to a few hundreds at higher Reynolds values \cite{tang2020, ren2021, holm2020}. Between each interaction with the environment, an interpolation scheme is usually exploited to avoid abrupt control changes in the environment, which could cause numerical instabilities. The considered papers either use a simple linear interpolation (as in \cite{tang2020}), or an exponential decay, as in \cite{rabault2019}.

Finally, the agent architecture networks are also consistent between the different studies, with two fully-connected layers used in all cases. While seven contributions appear to use pretty large layer sizes \cite{rabault2019, koizumi2018, tang2020, holm2020, paris2020, qinwang2021, li2021, yzwang2022}, smaller networks are successfully used for problems of similar dimensionalities \cite{tokarev2020, elhawary2020}. To the best knowledge of the authors, no large-scale study of the impact of the network architecture on the final agent performance was performed, and this choice remains most often empirical. Of particular interest is the use of a tailored single-step method in \cite{ghraieb2020}, that allows performing \emph{passive} control with extremely small networks (see section \ref{section:pbo}), which is because the agent is not required to learn a complex state-action relation, but only a transformation from a constant input state to a given action.

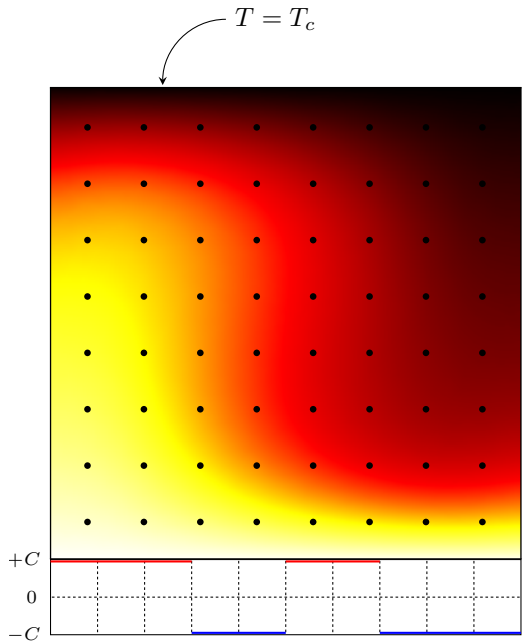
\begin{figure}
\centering
\def\sc{0.5}
\begin{tikzpicture}[	scale=\sc,
				probe/.style={circle, fill=black, inner sep=0pt, minimum size=2.5pt},
				empty/.style={ inner sep=0pt, outer sep=0pt},
				image/.style={anchor=south west,inner sep=0,scale=\sc, empty},
				grid/.style={dash pattern=on 1pt},
				hot/.style={red,thick},
				cold/.style={blue,thick}]
	
	\node[image] (image) at (0,0) {\fbox{\includegraphics[width=0.8\textwidth]{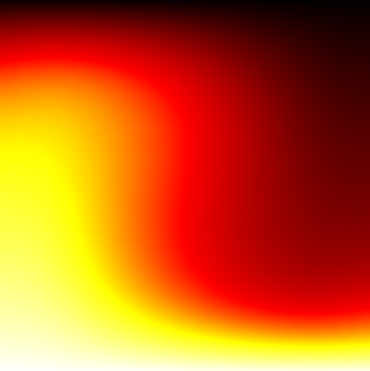}}};
	
	\foreach \x in {0,...,7}
    		\foreach \y in {0,...,7} 
			\node[probe] at (1 + 1.5*\x,1 + 1.5*\y) {};
			
	\node (top) at ($(image.north west) + (3,-0.2)$) {};
	\node[scale=1] (bc_top) at ($(top) + (3,2)$) {$T=T_c$};
	\draw[-stealth] (bc_top) to[out=180,in=90] (top);
	
	\node[empty] (botleft) at ($(image.south west) + (0.017,-2)$) {};
	\node[empty] (topleft) at ($(image.south west) + (0.017,0)$) {};
	\node[empty] (botright) at ($(image.south east) + (-0.017,-2)$) {};
	\node[empty] (topright) at ($(image.south east) + (-0.017,0)$) {};
	\node[empty] (midleft) at ($(botleft)!0.5!(topleft)$) {};
	\node[empty] (midright) at ($(botright)!0.5!(topright)$) {};
	
	\draw[fill=white] (botleft) rectangle (topright);
	\draw[grid] (midleft) -- (midright);
	
	\node[empty] (midtop1) at ($(topleft)!0.1!(topright)$) {};
	\node[empty] (midbot1) at ($(botleft)!0.1!(botright)$) {};
	\draw[grid] (midtop1) -- (midbot1);
	
	\node[empty] (midtop2) at ($(topleft)!0.2!(topright)$) {};
	\node[empty] (midbot2) at ($(botleft)!0.2!(botright)$) {};
	\draw[grid] (midtop2) -- (midbot2);
	
	\node[empty] (midtop3) at ($(topleft)!0.3!(topright)$) {};
	\node[empty] (midbot3) at ($(botleft)!0.3!(botright)$) {};
	\draw[grid] (midtop3) -- (midbot3);
	
	\node[empty] (midtop4) at ($(topleft)!0.4!(topright)$) {};
	\node[empty] (midbot4) at ($(botleft)!0.4!(botright)$) {};
	\draw[grid] (midtop4) -- (midbot4);
	
	\node[empty] (midtop5) at ($(topleft)!0.5!(topright)$) {};
	\node[empty] (midbot5) at ($(botleft)!0.5!(botright)$) {};
	\draw[grid] (midtop5) -- (midbot5);
	
	\node[empty] (midtop6) at ($(topleft)!0.6!(topright)$) {};
	\node[empty] (midbot6) at ($(botleft)!0.6!(botright)$) {};
	\draw[grid] (midtop6) -- (midbot6);
	
	\node[empty] (midtop7) at ($(topleft)!0.7!(topright)$) {};
	\node[empty] (midbot7) at ($(botleft)!0.7!(botright)$) {};
	\draw[grid] (midtop7) -- (midbot7);
	
	\node[empty] (midtop8) at ($(topleft)!0.8!(topright)$) {};
	\node[empty] (midbot8) at ($(botleft)!0.8!(botright)$) {};
	\draw[grid] (midtop8) -- (midbot8);
	
	\node[empty] (midtop9) at ($(topleft)!0.9!(topright)$) {};
	\node[empty] (midbot9) at ($(botleft)!0.9!(botright)$) {};
	\draw[grid] (midtop9) -- (midbot9);
	
	\draw[hot] ($(topleft) + (0,-0.05)$) -- ($(midtop3) + (0,-0.05)$);
	\draw[cold] ($(midbot3) + (0,0.05)$) -- ($(midbot5) + (0,0.05)$);
	\draw[hot] ($(midtop5) + (0,-0.05)$) -- ($(midtop7) + (0,-0.05)$);
	\draw[cold] ($(midbot7) + (0,0.05)$) -- ($(botright) + (0,0.05)$);
	\node[empty] at ($(botleft) + (-0.7,0.0)$) {\scriptsize $-C$};
	\node[empty] at ($(midleft) + (-0.5,0.0)$) {\scriptsize $0$};
	\node[empty] at ($(topleft) + (-0.7,0.0)$) {\scriptsize $+C$};
    
\end{tikzpicture}
\caption{\textbf{Illustration of the Rayleigh-B\'enard convection control setup} as presented in \cite{beintema2020}. The top wall temperature is set to a constant temperature $T=T_c$, while the bottom temperature profile is cut in 10 segments on which the temperature can take values equal to $T_h + C$ or $T_h - C$. On the left and right walls, adiabatic conditions are imposed. Finally, the temperature and velocity fields are collected on a grid of probes equispaced in the computational domain.}
\label{fig:rayleigh_benard_control}
\end{figure}

\subsection{Conjugate heat transfer}
\label{section:heat_transfers}

Although conjugate heat transfer systems governed by the coupled Navier--Stokes and heat equations seem natural candidates to extend the scope the DRL methodology and to increase the complexity of the targeted applications, the field has received little initial attention from the community. However, it may be starting gaining ground with two studies of DRL-based thermal control over the past two years.

In \cite{beintema2020}, the authors consider the closed-loop control of natural convection in a 2-D Rayleigh-B\'enard convection cell simulated with an in-house lattice-Boltzmann code at Rayleigh numbers (based on the time-averaged temperature difference between the upper and lower plates) ranging from $\Ra=10^3$ (just before the onset of convection) to $10^7$ (mild turbulence).
The set-up, synthesized in figure \ref{fig:rayleigh_benard_control}, is as follows: the upper plate and the time-averaged lower plate temperature distributions are assumed constant. A discrete PPO agent whose implementation relies on OpenAI's stable-baselines \cite{stable-baselines} is then used to provide (after a normalization step) a zero-mean, piecewise-constant lower temperature fluctuation with the intent to reduce the convective effects. The actor is a fully-connected neural network with two hidden layers of width 64, and the instantaneous reward is defined as the opposite of the instantaneous Nusselt number $\Nu$, which spurs the agent to minimize the convective effects at play. As in drag reduction applications, an array of probes is uniformy distributed over the computational domain to collect observations, under the form of both the temperature and velocity fields. 

A particularity of this implementation is the systematic use of the four most recent observations in the state buffer passed to the agent, an approach similar to that of \cite{koizumi2018}, although the authors do not provide insights about the impact of this choice on the agent performance. The agent is able to entirely stabilize the convective flow up to $\Ra=10^5$, and consistently outperform state-of-the-art linear approaches (proportional and proportional-derivative controllers) up to $\Ra=10^7$. Finally, the authors also illustrate the controllability limits of the system using the simplified Lorenz attractor system. By introducing a tunable artificial delay in the control, they show that exceeding half the Lyapunov time in delay results in a highly degraded performance of the learned control.

\begin{figure}
\centering
\fbox{\includegraphics[width=.7\linewidth]{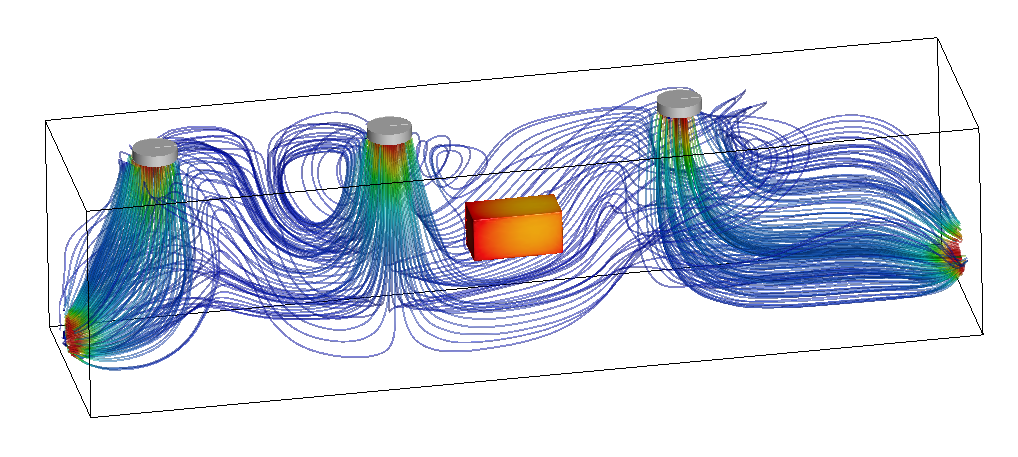}}
\caption{\textbf{Passive control of 3D forced convection in the context of workpiece cooling, reproduced from \cite{hachem2020}.} The positions of the three injectors are optimised in order to minimize the local temperature gradients in the workpiece during the cooling process.}
\label{fig:workpiece_cooling}
\end{figure}

Passive control of a similar Rayleigh-B\'enard natural convection problem is performed in \cite{hachem2020} with the single-step approach presented in section \ref{section:pbo}. Compared to \cite{beintema2020}, the authors report excellent control efficiency using much smaller networks (two hidden layers of width 2 vs. 64) and less parallel environments (8 vs. 512) at $R_a = 10^4$, a value for which the optimal control determined in \cite{beintema2020} ends up being actually time-independent (unlike at higher Rayleigh numbers). The authors then use the same approach and network architecture to minimize open-loop the inhomogeneity of temperature gradients across the surface of two and three-dimensional hot workpieces under impingement cooling in a closed cavity, identifying either optimal positions for cold air injectors relative to a fixed workpiece position, or optimal workpiece position relative to a fixed injector distribution (see illustration in figure \ref{fig:workpiece_cooling}).

\subsection{Shape optimization}
 
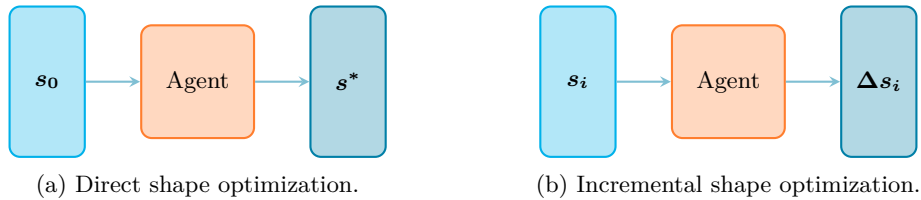
\begin{figure}[h!]
\centering
\begin{subfigure}[t]{.4\textwidth}
	\centering
	\begin{tikzpicture}[arrow/.style={thick,color=bluegray3,rounded corners}]

		\node[	rectangle, rounded corners, draw=blue1, 
				fill=blue4, thick, minimum width=1cm, 
				minimum height=2cm, text centered,
				align=center] 
				(x) at (0,0) {\footnotesize $\bm{s_0}$};
			
		\node[	rectangle, rounded corners, draw=orange1, 
				fill=orange4, thick, minimum width=1.5cm, 
				minimum height=1.5cm, text centered] 
				(NN) at (2,0) {\footnotesize Agent};
			
		\node[	rectangle, rounded corners, draw=bluegray1, 
				fill=bluegray4, thick, minimum width=1cm, 
				minimum height=2cm, text centered,
				align=center] 
				(y) at (4,0) {\footnotesize $\bm{s^*}$};
			
		\draw[-stealth,arrow] (x.east) -- (NN.west);
		\draw[-stealth,arrow] (NN.east) -- (y.west);
	
	\end{tikzpicture}
	\caption{Direct shape optimization.}
	\label{fig:shape_optimization_direct}
\end{subfigure} \qquad
\begin{subfigure}[t]{.4\textwidth}
	\centering
	\begin{tikzpicture}[arrow/.style={thick,color=bluegray3,rounded corners}]

		\node[	rectangle, rounded corners, draw=blue1, 
				fill=blue4, thick, minimum width=1cm, 
				minimum height=2cm, text centered,
				align=center] 
				(x) at (0,0) {\footnotesize $\bm{s_i}$};
			
		\node[	rectangle, rounded corners, draw=orange1, 
				fill=orange4, thick, minimum width=1.5cm, 
				minimum height=1.5cm, text centered] 
				(NN) at (2,0) {\footnotesize Agent};
			
		\node[	rectangle, rounded corners, draw=bluegray1, 
				fill=bluegray4, thick, minimum width=1cm, 
				minimum height=2cm, text centered,
				align=center] 
				(y) at (4,0) {\footnotesize $\bm{\Delta s_i}$};
			
		\draw[-stealth,arrow] (x.east) -- (NN.west);
		\draw[-stealth,arrow] (NN.east) -- (y.west);
	
	\end{tikzpicture}
	\caption{Incremental shape optimization.}
	\label{fig:shape_optimization_incremental}
\end{subfigure}
\caption{\textbf{Direct and incremental DRL-based shape optimization techniques} present in the literature. In direct shape optimization (left), the agent is used as a proxy to optimize a direct mapping from a constant, initial state vector $\bm{s_0}$ to the optimal state $\bm{s^*}$, using a degenerate, single-step DRL algorithm. In incremental shape optimization (right), the agent learns the adequate mapping from the current state vector $\bm{s_i}$ to an incremental modification to apply to the latter, hence determining a path of incremental deformations to apply from $\bm{s_0}$ to $\bm{s^*}$. In both cases, multiple episodes are required for the agent to converge.}
\label{fig:shape_opt}
\end{figure}

Shape optimization is another field fundamentally interrelated with flow control, that can seem as a natural domain application for the DRL techniques covered above. Nonetheless, it is worth noticing that shape optimization generally consists in determining a fixed shape meeting a set of required criteria (\eg high lift-to-drag ratio, low pressure loss). 
This is not \textit{per se} the original purpose of DRL, that aims at identifying optimal state-to-action relations (by means of neural network training) and is thus best suited to dynamically manipulate a deformable shape.
Two approaches exist in the literature in the context of DRL-based shape optimization, a first one that directly optimizes state-independent shape parameters (hence, \textit{direct shape optimization} \cite{viquerat2021}) and a second one that incrementally modifies an initial shape into an optimal one (hence, \textit{incremental shape optimization} \cite{yan2019, li2020, qin2021}). The conceptual differences between these two approaches are illustrated in figure \ref{fig:shape_opt}, and their implementations are detailed in the following paragraphs.\footnote{Although not directly included in the scope of the current review, it is worth mentioning the work from Lampton \textit{et al.\@}  \cite{lampton2008}, who considered the use of standard Q-learning method for shape optimization in 2008. In this contribution, optimization of airfoil geometries with four free parameters is considered, and the optimal policy is obtained by updating a Q-table in a temporal-difference fashion.}

\begin{figure}
\centering
\begin{subfigure}[t]{.25\textwidth}
	\centering
	\fbox{\includegraphics[width=\linewidth]{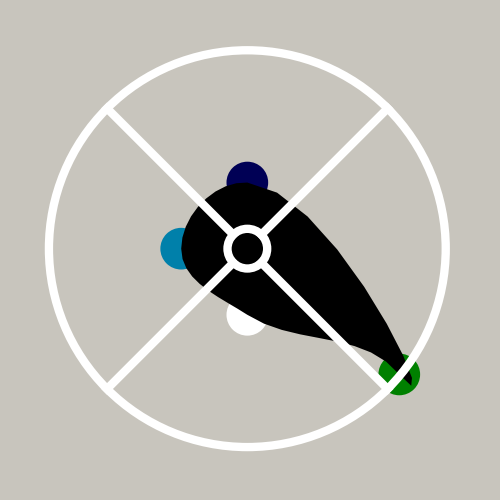}} 
	\caption*{}
\end{subfigure} \qquad
\begin{subfigure}[t]{.25\textwidth}
	\centering
	\fbox{\includegraphics[width=\linewidth]{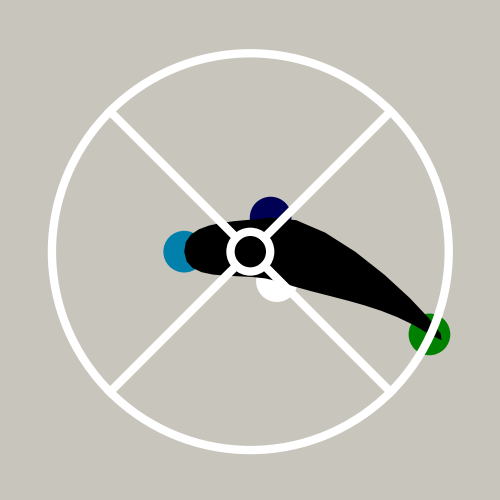}} 
	\caption*{}
\end{subfigure} \qquad
\begin{subfigure}[t]{.25\textwidth}
	\centering
	\fbox{\includegraphics[width=\linewidth]{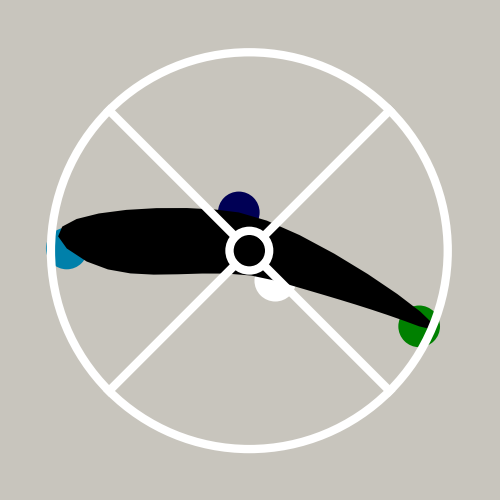}} 
	\caption*{}
\end{subfigure}
\caption{\textbf{Shapes of optimal lift-to-drag ratio obtained with direct shape optimization} with 3 free parameters (left), 9 free parameters (center) and 12 free parameters (right), reproduced from \cite{viquerat2021}. The shape parameterization relies on B\'ezier curves joining control points, the agent controlling their position and local curvature radius. These shapes were obtained by learning an optimal mapping from a simple cylinder.}
\label{fig:direct_shape_opt}
\end{figure}

In direct shape optimization, the agent is used as a proxy to optimize a direct mapping from a constant, initial state vector $\bm{s_0}$ to the optimal state $\bm{s^*}$. This approach is implemented in \cite{viquerat2021} using single-step DRL (the degenerate class of DRL algorithms intended to optimize state-independent agent behavior) to design 2-D aerodynamic profiles without any \textit{a priori} knowledge, feeding systematically an initial circle as input to the agent in single-step episodes (hence the adjective \emph{stateless}). In practice, the shapes are described by a set of B\'ezier curves connecting the same number of control points, each with 3 free parameters (2 coordinates, plus a local curvature radius). As shown in figure \ref{fig:direct_shape_opt}, the agent is able to design airfoil-like shapes maximizing the lift-to-drag ratio at Reynolds numbers of about a few hundred, which takes between one and three thousand CFD evaluations (\textit{i.e.} single-step episodes) for problem dimensionality ranging from 3 to 12, respectively. 

The literature proposes three other DRL-based shape optimization contributions conversely relying on incremental shape transformations, with the incremental modifications in \cite{yan2019,qin2021} taking the current geometric parameters as input (of dimension 8 and 10, respectively), while the input states in \cite{li2020} consist in a distribution of wall Mach number (of dimension 4). In all three contributions, a pre-trained surrogate or a simplified model is used, either for full agent training, or to perform an initial learning phase before re-training on a CFD environment using transfer learning. A key difference lies in the fact that the authors in \cite{yan2019,qin2021} always use the same input state and consequently produce a single optimized shape per training, while \cite{li2020} relies on a set of input states randomly selected at the beginning of each episode, meaning that the trained agent can be successfully re-used in production on out-of-training input shapes.

From an algorithmic point of view, the choices are in line with those reported in the previous sections. All algorithms are actor-critic, either PPO \cite{li2020} or DDPG \cite{yan2019, qin2021}, using fully-connected networks with 2 or 3 layers of width from 200 to 512 neurons per layer, except for \cite{viquerat2021}. As it represents an arbitrary design choice in this specific application, the number of steps per episode is low, ranging from 5 to 20. A simple reward signal based on the lift-to-drag ratio is used in \cite{yan2019} and \cite{viquerat2021}, but more complex designs were used in the other two contributions. In \cite{li2020}, the instantaneous reward is based on the difference of drag between the current and the previous generation, while \cite{qin2021} exploits a complex reward expression based on the results of a principal component analysis. Overall, it is extremely difficult to draw conclusions from these different approaches. Moving forward, a careful performance comparison between direct and incremental approaches constitutes a topic of outmost importance, but a more specific study focused on reward design could also be of great practical interest. 

\subsection{Swimming}
\label{section:swimming}

The control of swimmers has been a pioneering field for applying  deep reinforcement learning to fluid mechanics problems, with a couple of contributions \cite{novati2017, verma2018} building on early seminal studies focusing RL for schooling \cite{gazzola2014, gazzola2016}. In \cite{novati2017}, the kinematics of two swimmers in a leader-follower configuration are analyzed based on 2-D simulations of viscous incompressible flows.
The first fish (leader) swims with a steady gait and the second fish (follower) uses DRL to adapt its behaviour dynamically to account for the effects of the wake encountered. The retained algorithm is DQN (see section \ref{section:dqn}), with input states made up of the lateral displacements and orientation of the follower compared to leader, as well as the two most recent actions, and the tail-beat status. An $\epsilon$-greedy strategy is used to perform exploration, with randomness decaying from 0.5 to 0.1 over the course of learning. The reward design is straightforward, and increasingly penalizes the follower when it strays too far away from the leader path:

\begin{equation}
\label{eq:reward_swimming}
	r_t = 1 - 2 \frac{\left| \Delta y \right|}{L},
\end{equation}

where $\Delta y$ is the aforementioned deviation, and $L$ is the length of the swimmer. It takes roughly \num{100000} transitions to learn the optimal behavior, and the results indicate that swimming in synchronized tandem (with the follower seeking to maintain its position in the center of the leader's wake, and its head synchronized with the vortices shed by the leader) can yield up to about 30\% reduction in energy expenditure for the follower.

\begin{figure}
\centering
\fbox{\includegraphics[width=.6\linewidth]{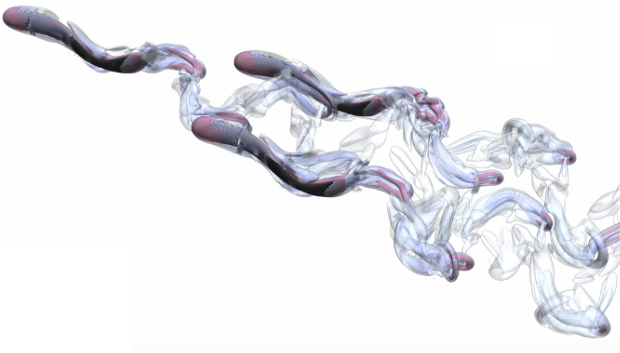}}
\caption{\textbf{Coordinated schooling of three swimmers, reproduced from \cite{verma2018}.} The two followers interact with both rows of the wake shedding to increase their swimming efficiency.}
\label{fig:schooling}
\end{figure}

Reference \cite{verma2018} is a follow-up of \cite{novati2017} extended to 3-D schooling configurations, as illustrated in figure \ref{fig:schooling}. A key contribution of this study is the use of a recurrent neural network, as the authors advertise (and demonstrate by providing performance comparisons with standard feedforward neural network) a greatly accelerated learning process using long-short term memory (LSTM) cells to encode the unsteadiness of the value function, which in turn is found to enable far more robust smart-swimmers. The retained recurrent network is composed of three layers of fully-connected LSTM units. The DQN algorithm with Adam optimizer is used to perform training in a temporal-difference manner, using again an $\epsilon$-greedy exploration, with randomness decaying from 1 to 0.1. The training procedure requires \num{46000} transitions (a reduction by roughly 50\% with respect to the LSTM-less 2-D case). The results support the conjecture that swimming in formation is energetically advantageous, with the trained fishes showing collective energy-savings behaviors by appropriately placing themselves in appropriate locations in the wake of other swimmers and interacting judiciously with their shed vortices. An almost identical set-up (\textit{i.e.} DQN algorithm exploiting an LSTM-based agent) is used in \cite{zhu2021}, to tackle a series of different swimming problems, namely (i) point-to-point travel in quiescent flow, with reward based on normalized distance to target, (ii) holding a steady position in a rotating fluid flow, with reward based on averaged translation velocity of the fish center of mass, and (iii) holding a steady position in a Karman vortex street. The authors also emphasize the necessity to provide richer information to the agent to reduce variability over multiple episodes. The retained approach consists in feeding the agent with informations about the fish dynamics over the last four periods (\eg, depending on the case, distance to objective, orientation of the swimmer, mean swimming velocities) and to add the actions taken over the same history steps, which indeed is found to yield stable learning and efficient swimming strategies. A similar setup (DQN with LSTM units) is employed in \cite{zhu2022}, coupled to an LBM solver. In this latter paper, a swimmer learns to reach a given destination located upstream of its position in a vortical flow. \blue{Two other contributions are to be noticed \cite{yan2020, yan2021}, in which the authors consider the coupling of an actor-critic agent with a strongly-coupled fluid-solid interaction solver, based on the arbitrary lagrangian-eulerian method. In this context, the swimmer learns to perform several tasks, such as swimming along pre-defined curvilinear trajectories, or learning to avoid obstacles while reaching a target position.}

\subsection{Microfluidics}
\label{section:microfluidics}

Micro-fluidics is one of the first fluid dynamics problems tackled with deep reinforcement learning techniques, but the related literature has since stalled to a single contribution from 2019 \cite{lee2019} (along with an additional experimental study \cite{dressler2018} reviewed in section \ref{section:revexpe}). In \cite{lee2019}, the authors consider the inverse design problem of flow sculpting, in which a relevant sequence of micro-pillars is designed to controllably deform an initial flow field into a desired one. A double-DQN agent (section \ref{section:dqn}) is used that implements a convolutional policy, the full flow map being passed as input state \cite{ddqn}. The agent network is composed of three convolutional/max-pooling layers followed by three batch-norm/fully-connected layers. The DDQN is supplemented with an experience replay method \cite{hes}. The implemented reward is based on a pixel match rate (PMR) that measures the similarity of the current flow with the target flow. This contribution also contains an interesting analysis comparing DDQN performance with that of canonical methods, \eg, genetic algorithms and brute force approaches.

\subsection{Other applications}
 
This section connects to other contributions of the literature applying DRL to more restricted \blue{sub-domains of fluid mechanics}, \eg, turbulence model generation \cite{novati2021}, sloshing suppression \cite{xie2021} or instability mitigation in fluids \cite{belus2019}, among others. It is worth insisting that the scarcity of publications on these topics does not reflect a lack of interest or priority, but rather the suddenness with which DRL has opened up new opportunities for a wide range of applications, as was already clear from the previous sections. \blue{Note also, the literature features a few other publications pertaining to more peripheral domains of applications, \eg energy efficiency~\cite{kazm18,zhang19} or wave energy converters \cite{anderlini2018, bruzzone2020}. These generally use 
low-dimensional models basically unrelated to the equations of fluid dynamics, and are thus not formally considered to keep the scope of the review well-defined.}

\subsubsection{Flow control}

\begin{figure}
\centering
\fbox{\includegraphics[width=.95\linewidth]{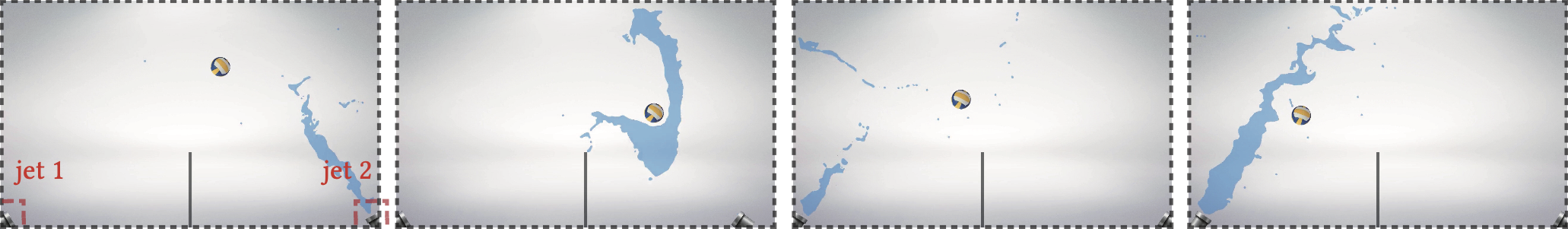}}
\caption{\textbf{Rigid body control using steerable fluid jets, reproduced from \cite{ma2018}.} Here, the goal is to push the ball back and forth from one jet to another.}
\label{fig:fsi}
\end{figure}

In an early contribution by Ma \textit{et al.\@} in 2018 \cite{ma2018}, a TRPO (section \ref{section:ppo}) agent learns to play different games (from rigid body balancing to complex music-playing games) based on the control of rigid body by steerable fluid jets, as illustrated in figure \ref{fig:fsi}. Regarding the environment, the Navier--Stokes equations are marched in time using a grid-based fluid-solid solver with adaptive refinement. A convolutional auto-encoder trained on-the-fly is used to efficiently extract fluid flow features from the environment. After their dimensionality has been reduced to an acceptable range, those are combined with rigid body features and serve as input for the agent, which is shown to significantly improve the learning speed compared to using rigid body features only. The state vector, whose size is lower than 100 elements, is fed to a standard fully-connected network of size $[128,64,64,32]$, which yields typical training times in a range from 2 to 20 hours, depending of the game played.

\begin{figure}
\centering
\fbox{\includegraphics[width=.6\linewidth]{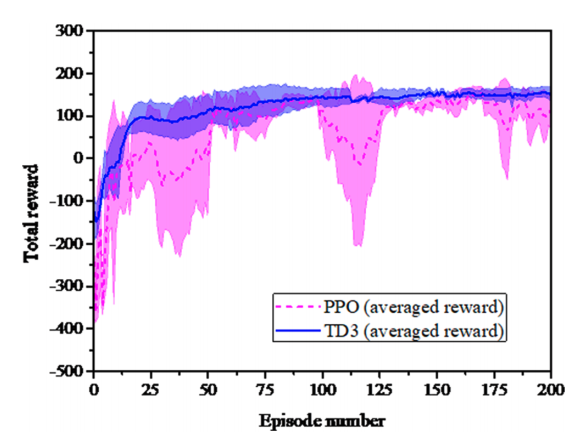}}
\caption{\textbf{Performance comparison of proximal policy optimization (PPO) and twin-delayed deep deterministic policy gradient (TD3) for sloshing suppression task, reproduced from \cite{xie2021}.} Although similar performance levels are obtained, the learning process proves to be more stable for TD3.}
\label{fig:sloshing_ppo_td3}
\end{figure}

Another under-represented type of application is the control of sloshing in tanks, despite obvious practical interest for engineering applications, such as liquid carriers in ground, marine, or air transport vehicles, as well as in earthquake excited water supply towers. Reference \cite{xie2021} is the only contribution in the field, that considers suppressing sloshing in a tank initially submitted to a sinusoidal excitation using two active controlled horizontal baffles. The comparison of two policy-gradient algorithms, namely PPO (section \ref{section:ppo}) and TD3 (section \ref{section:td3}) is a key contribution of this study. The state information consists of the positions of the baffles, as well as the elevation and vertical velocities of two additional probes in the tank. 
Given such inputs, the agent provides in return the horizontal velocities to be applied to the baffles. For both algorithms, the actor is composed of a fully-connected network with two layers of width 64. Actions are taken by the agent every 30 numerical time-steps, one episode consisting in 200 actions, linearly interpolated from one time-step to the following. The reward is equal to the time-averaged sloshing height, plus a penalization term to limit the displacements of the baffles. Good convergence is reported both for PPO and TD3, although learning proves to be more stable using TD3, as shown in figure \ref{fig:sloshing_ppo_td3}. With direct learning, the authors notice a lack of robustness when applying the learned strategy beyond the largest time used during training, which they show can be overcome using behavior cloning to pre-train the agent.

\begin{figure}
\centering
\begin{subfigure}[t]{.45\textwidth}
	\centering
	\fbox{\includegraphics[width=\linewidth]{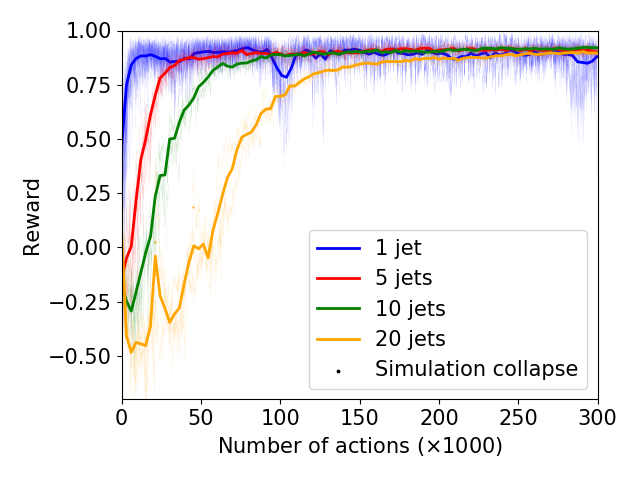}} 
	\caption{Naïve training method}
\end{subfigure} \qquad
\begin{subfigure}[t]{.45\textwidth}
	\centering
	\fbox{\includegraphics[width=\linewidth]{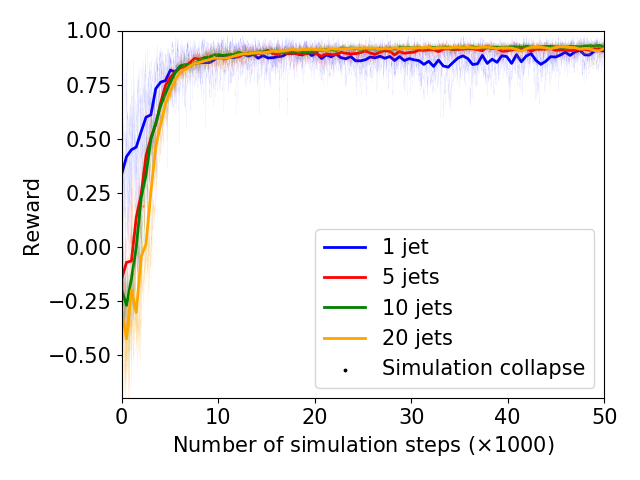}} 
	\caption{Training approach based on invariance}
\end{subfigure}
\caption{\textbf{Learning curves obtained with the naïve learning technique (left) and with the invariance-based approach (right), reproduced from \cite{belus2019}}. On the left figure, it can be seen that increasing the number of jets significantly increases learning time, here counted in number of actions. On the right figure, similar training times (counted in simulation steps) are required, whatever the number of jets (and therefore the action space dimension) used to control the instability.}
\label{fig:fluid_film}
\end{figure}

Another noteworthy contribution is that of Belus \textit{et al.\@} \cite{belus2019}, that introduces a technique based on invariants intended for problems with large dimensional (up to 20) actions spaces. In this study, a PPO agent (section \ref{section:ppo}) is used to mitigate the natural instabilities developping in a 1-D falling liquid film using small jets blowing orthogonally to the flow direction. The number of jets and their positions can vary, leading to different levels in control complexity. A three-layer, fully-connected network of size $[128,64,64]$ is used, with actions provided every 50 numerical time-steps to a variable number of jets, based on local inputs recorded in the vicinity of each jet. Finally, the reward function steers the agent to alleviate the waves arising from the instability of the flow. Three training methods are compared, that differ by their ability to handle a large number of control jets: (i) local states are concatenated and flattened before being fed to the actor, its output dimensionality being equal to the number of jets; (ii) a similar approach is used, but instead of being flattened, the input states are fed as is to a convolutional network; (iii) the vicinity of each jet is considered a local environment and used to provide some states and a reward to a unique agent. This latter approach relies on the translational invariance of the physical problem. It significantly enhances the amount of experience collected by the agent during an episode, which the authors show allows tackling large dimensional action spaces without increasing the amount of simulation time, as shown in figure \ref{fig:fluid_film}.

Vibration suppression was also considered in the context of an academic test-case \cite{zheng2021}. In this article, the authors consider the reduction of the vibration of a cylinder in a $Re=100$ flow, its motion being constrained by a damping device and a spring. Similarly to the drag reduction cases, the cylinder is equipped with synthetic jets, which mass flow rate is controlled by the agent. The amplitude of the vibrations is successfully reduced by more than 80\% using a soft actor-critic (SAC) agent, and these results are compared with those of an active learning approach.

\subsubsection{Turbulence modeling}

An approach somehow similar to that in \cite{belus2019} is used in another study by Novati \textit{et al.\@} \cite{novati2021} to adjust the coefficients of an eddy viscosity closure model in the attempt to reproduce the energy spectrum of DNS computations. To this end, multiple agents are dispatched in the computational domain, with each agent controlling locally the dissipation coefficient of the Smagorinsky SGS model. The provided states are a blend of local (invariants of the gradient and Hessian of the velocity field) and global quantities (modes of the energy spectrum, rate of viscous dissipation, total dissipation). Two types of reward are proposed, based either on the Germano identity, or on a distance to a pre-computed DNS spectrum. The agent uses the remember-and-forget experience replay method, in which networks update are performed within a trust region, using a buffer holding the most recent transitions collected by the policy (which supposedly greatly improves the sample efficiency by enabling data to be reused multiple times for training) while dismissing those actions too unlikely under the current policy. The network parameters are shared between agents, and their aggregated experiences are collected in a shared dataset used for training. This approach was extended by \cite{kim2022} in order to build SGS models for wall-bounded turbulence, by introducing physical constraints in the training of a TD3 agent.

\subsubsection{\blue{Differential equation resolution}}

\blue{An original contribution from Wei \textit{et al.} \cite{wei2019} is concerned with the direct resolution of ordinary and partial differential equations by exploiting a modified actor-critic framework, and especially the resolution of the Navier-Stokes equations. In this paper, the authors rely on the actor to propose candidate solutions to the considered equation, while the "critic" is in fact returning the residual of the candidate solution when injected in the governing equations of the problem. This approach shares similar traits both with standard actor-critic techniques, due to the way candidate solutions are generated, and with unsupervised physics-informed methods, due to the substitution of the critic with a computation of the residual of the candidate solutions. Excellent agreement is observed on several equations, such as the Schr\"odinger equation or the Navier-Stokes equations, among others.}

\section{\blue{Deep reinforcement learning} and experimental fluid dynamics}
\label{section:revexpe}

The coupling of DRL and experimental fluid mechanics remains insufficiently explored, with only 3 out of the 32 papers compiled in this review applying DRL for experimental flow control purposes. Besides the possibly limited access to experimental devices for DRL practitionners, this is likely because several challenges such as controllability (the ability to efficiently reach a given state), observability (the ability to reliably measure changes in the state), sensitivity (to noise and system uncertainty) and system delays (see section \ref{section:challenges}) become increasingly important in experimental setups, even though they have received little attention in the context of idealized numerical environments.

\subsection{Drag reduction}
\label{section:dragexpe}

In \cite{fan2020}, a drag reduction problem similar to that in figure \ref{fig:drag_reduction_downstream_rotating} is considered, where an agent is given control over the angular velocity of two rotating cylinders located in the wake of a fixed principal cylinder. The Reynolds number is about $\Re=10^4$, and the agent is allowed to interact with the environment every 0.1 s. An entire episode last 40 s, plus additional time consumption for initialisation (4s, the time needed to wash out the transient before collecting any data) and for the reset procedure (2 mn, the time needed for the entire system to come back to rest). Overall, an experimental episode lasts between 3 and 4 mn. A TD3 agent (section \ref{section:td3}) based on Tensorflow is used, the updates being performed only between episodes with a reward function similar to (\ref{eq:reward_drag_lift}). The states provided to the agent are the drag and lift coefficients measured on the main cylinder and the two control cylinders (an approach noticeably different from that described in section \ref{section:drag_reduction}). A key outcome of this study is the necessity to high pass filter the experimental states before they are fed to the agent, as a comparison of the performance with and without providing beforehand the experimental states as input to a Kalman filter shows that the agent is essentially unable to learn an efficient strategy without the filtering stage. Additional experiments are also performed to account for the power loss due to the friction of the control cylinders

\subsection{Flow separation}

In \cite{shimomura2020}, a DQN agent (section \ref{section:dqn}) learns to perform flow reattachment behind a NACA~0015 airfoil by controlling the burst frequency of a plasma actuator at $\Re = \num{6.3e4}$. Two different angles of attack are considered, namely 12$^\circ$ and 15$^\circ$. The states provided to the agent consist of the unfiltered time-series data of the pressure at the surface of the airfoil, recorded through a set of 29 holes with high-frequency sensors, eventually downsampled to a total of 80 values. The actions are selected among a set of pre-defined burst frequencies, that includes four different values as well as an "off" choice. The reward is zero if the flow is not attached, and one if it is attached, as determined from the pressure coefficient at the trailing edge of the airfoil. The DQN agent achieves a satisfactory learning at the first angle of attack of 12$^\circ$, with efficient strategies available after as little as 200 episodes, although not more efficient that a naive open-loop control with adequately selected burst frequency. Conversely, the agent significantly outperforms the naive open-loop design at the second angle of 15$^\circ$, but learning is then much more challenging and takes about up to 800 episodes.

\subsection{Microfluidics}

The problem considered in \cite{dressler2018} relates to the performance of microfluidics experiment platforms when operated on extended periods of time. To overcome degraded flow stability beyond a certain timescale, the authors introduce a DRL agent to adjust the flow conditions and maintain the experiment operability \emph{in an experimental device}. Two low-Reynolds applications are considered, namely the positioning of an interface between two miscible flows, and the dynamic control of the size of water-in-oil droplets within a segmented flow. On both applications, the performances of a DQN \cite{dqn} agent and a model-free episodic control (MFEC) \cite{mfec} are compared, although it must be noted that the algorithm run with different interaction frequencies (250 actions per episode for DQN, vs. 150 for MFEC) due to equipment limitations. Observations are obtained from a high-speed camera and processed into an $84 \times 84$ pixels frame. In the first experiment, the reward is obtained calculating the distance between the current observed interface and its target position, while in the second experiment it is computed from the estimated radii of the generated droplets. 
The authors find that DQN requires a considerable amount of frames (approximately \num{145000} in the first experiment) to surpass human-level performance, albeit with large-scale fluctuations, while MFEC requires a reasonable number of frames to improve and reach a stable level of performance (approximately \num{11000} frames in the first experiment), but does not reach the peak performance of DQN in the first case.

\section{Transversal remarks}\label{section:transverse}

The contents of previous sections, although presented per application, helps identify trends regarding several technical aspects of state-of-the-art contributions in DRL for fluid flow problems. The present section underlines some of the latter, and raises open questions regarding possible future improvements in the field.

\subsection{Taking on the challenges}

Based on this review, we deem there is a good understanding of the key issues relevant to fluid flow problems. Many of the compiled references are primarily aimed at proving either feasibility in such or such sub-domain, or beyond state-of-the-art performance of such or such algorithm, but several milestone contributions assess the ability of novel developments to 
increase the complexity of the problems presented to the DRL agent. 
Among the challenges listed in section~\ref{section:challenges}, computational efficiency \cite{rabaultkuhnle2019}, stochasticity \cite{ghraieb2020,beintema2020,ren2021} and partial observability \cite{paris2020} have received the most attention, but robustness and delays remain largely ignored (save for the unique combination of stochasticity and post-action time delays examined in \cite{beintema2020}), even though real-world environments likely feature all mechanisms in strong interaction one with another. \green{In particular, \cite{mao2022} demonstrates the necessity of including both a physics-informed delay and regressive models in the Markov decision process (not just one or the other) to achieve a robust and temporal-coherent control under weak turbulent conditions.} 

A lot has been achieved in a short period of time, but many related issues remain to be addressed for which the RL literature provides a number of a off-the-shelf methods already proved fruitful in different context (mostly robotics), that could help reach even higher levels of performance and robustness. Typical examples include learning a model of the environment in such a way that errors in the model do not degrade the asymptotic performance \cite{chua2018,buckman2018}, or wrapping redundant states into equivalent classes of canonical spaces \cite{wu2017} to increase the data efficiency; 
using data augmentation and randomization techniques to train over a wide distribution of states \cite{tobin2017,lee2019net} or partitioning the initial state distribution and training different policies later to be merged \cite{ghosh2017} to alleviate stochasticity; optimizing for worst case expected return objectives \cite{mankowitz2019} or pursuing soft-robustness \cite{derman2018} to improve robustness; \red{adding incentives to increase the policy entropy to provide more choices in solving a problem when situations are changed from the training, and thus to ease transfer learning~\cite{eysenbach2021maximum}};
using the frameworks of partially observable Markov decision process \cite{cassandra1994} and delay-aware Markov Decision Process \cite{chen2021} to account for partial observability and delayed dynamics.

\subsection{Providing guidelines for the selection of the \blue{deep reinforcement learning} algorithm}

An obvious preference for policy gradient techniques appears from the review, with PPO the clear-cut go-to algorithm; see table \ref{table:method_usage}. This is noteworthy because PPO is an on-policy algorithm, that updates the policy used to generate the training data (in contrast to off-policy algorithms, that also learn from data generated with other policies). PPO is generally acknowledged to improve the sample efficiency of regular actor-critic techniques, but there could be a fad component to this rise to prominence (partly attributable to the early open-source code release of several projects relying on this technique \cite{rabault2019, rabaultkuhnle2019, tang2020}), given that off-policy methods are expected to have even higher sample efficiency, and that most authors fail to explain the rationale for choosing a particular algorithm over another. Given the high CPU requirements of CFD solvers (that remains an important limitation regarding the application of DRL to 3-D flows of engineering importance), this calls for more careful, consistent and systematic testing of state-of-the-art on- and off-policy techniques in a fluid mechanics context. At the time of writing, only two such comparison studies are available in the literature, namely PPO vs. TD3 in \cite{xie2021}, and DQN vs. MFEC in \cite{dressler2018}.

\begin{table}
\caption{\textbf{Usage frequency of different \blue{deep reinforcement learning} algorithms in the articles considered in the present review.} \blue{Proximal policy optimization} is obviously the most spread method, most probably due to several open-source releases. 
}
\label{table:method_usage}
\centering
\medskip

{
\setlength{\tabcolsep}{4pt}
\begin{tabular}{rr}																																			\toprule
\red{Deep Q-networks (DQN)} & 5\\
\red{Double deep Q-networks (DDQN)} & 2\\
\red{Advantage actor-critic (A2C)} & \blue{4}\\
\red{Proximal policy optimization (PPO)} & \green{15}\\
\red{Trust-region policy optimization (TRPO)} & 1\\																					\red{Deep deterministic policy gradient (DDPG)} & 4\\
\red{Twin-delayed deep deterministic policy gradient (TD3)} & 3\\
\red{Soft actor-critic (SAC)} & 1\\
\red{Single-step PPO (PPO-1)/Policy-based optimization (PBO)} & 3\\
{Others} & 3\\											
								\bottomrule
\end{tabular}}
\end{table}

\subsection{Fighting the reproducibility crisis}

DRL a very fast-moving field, and as the number of contributions is growing, it becomes harder and harder to make a proper comparison between DRL algorithms, all the more so as a bevy of algorithms have been developed, to be used from dedicated libraries (\eg Tensorforce \cite{tensorforce}, Stable Baselines \cite{stable-baselines}, OpenAI Baselines \cite{baselines}) or implemented in-house (which relates to 10 out of the 32 reviewed contributions).
Compounding the matter are the high amount of time needed to train DRL agents, that creates a high barrier for reevaluation
of previous work,; the general lack of complete information regarding the network architecture (\eg size and depth of the hidden layers, activation functions, normalization, initialization) and training procedure (\eg optimizer, batch size, number of epoch per update, update frequency, learning rate); and (for numerical environments) the additional variance in the numerical solutions themselves.

Encouraging the open sourcing of appropriate code on public git repositories is thus a critical step to ensure the reproducibility and durability of the developments, to maximize their impact, and to ultimately help establish DRL as a mature and stable technique for the analysis and design of complex flow systems. In this respect, it is disappointing to note
that only 9 out of the 32 studies compiled in this review have come with such open-source releases \cite{rabault2019, tokarev2020, tang2020, elhawary2020, holm2020, rabaultkuhnle2019, viquerat2021, novati2021, belus2019}.
Creating and providing exhaustive benchmark datasets and metrics is another alternative that would certainly add value to the community, and lay the ground for solid further developments in the field. The authors take this opportunity to underline the work by Wang \etal \cite{qwang2022}, that proposes a specific, ready-to-use platform for the coupling of DRL with CFD applications. This platform uses the Tensorforce library for the control side, and OpenFoam for the numerical solver.

\subsection{Other research gaps}

The present review has also allowed us to identify several other important gaps to consider when evaluating the progress of DRL for practically meaningful fluid mechanics.

\textit{$\circ$ Network architecture:} almost all provided references use fully-connected networks, with two or three hidden layers, each holding a number of neurons in the range from a few tens to a few hundreds. Nonetheless, our review did not reveal any large-scale study of the impact of the network architecture on the agent performance, and the choice remains most often empirical. The single-step method used in \cite{ghraieb2020} is especially interesting in this regards, as it succeeds in learning optimal state-independent policies from extremely small networks. It should also be noted that the successful use of long-short term memory cells instead of regular fully-connected networks was advertised in swimming applications \cite{verma2018, zhu2021}, and that additional comparative experiments on different problems could lead to a more systematic use of such architectures.

\textit{$\circ$ State space dimensionality:} in some cases, state selection seems arbitrary, which can lead to either (i) incomplete observations or (ii) a too large inputs to the actor, which can be detrimental to learning. Specific methods have been proposed to tackle this issue, either by adding an intelligent state selection mechanism \cite{paris2020}, or by exploiting state compression \cite{ma2018}. Shall they be pursued further, such efforts could lead to systematic techniques for state input from CFD environments.

\textit{$\circ$ Action space dimensionality:} in most contributions, the dimension of the action space remained limited, usually between 1 and 3. In this context, Belus \textit{et al.} showed that exploiting the physical invariants of the problem was a particularly efficient way to tackle action spaces of larger dimensions (up to 20) \cite{belus2019}. 

\textit{$\circ$ Time granularity:} the frequency at which the agent interacts with its environment is usually set based on physical considerations, but the ratio of the typical physical time scale to the action time-step remains highly variable from one contribution to another (even for very similar cases; see table \ref{table:drag_reduction_summary}). Since this hyper-parameter can dramatically affect the attainable performance of the agent and the difficulty of the learning task (too large intervals lead to inefficient actions, while too small intervals hinder the learning process), the development of systematic selection criteria is another aspect that could benefit the community and help close the gap with real-world testing.

\textit{$\circ$ Scale effects:} \red{due to the scarce literature that considers the application of DRL to experimental flow systems, the existence of potential scale effects when pivoting from idealized numerical systems to real physical models has never been assessed at the time of writing. Such effects arising from imperfect numerical modeling could lead to considerable deviation when model control laws are extrapolated to prototype values, which can easily, \eg, impact wave dynamics or optimal propeller design.}

\textit{$\circ$ \blue{Comparison with other control methods}:} \red{it is generally acknowledged that the main advantage for using DRL over more traditional control or optimization algorithms lies in its ability to reveal complex and unanticipated solutions or parameter relations, where most control strategies used in published works about active flow control rely on relatively simple harmonic or constant control input. That being said,} \blue{the general literature considering the comparison of DRL control approaches with other control methods is extremely scarce. In the context of the coupling with fluid dynamics, only two references could be found that include comparisons of DRL with other approaches. For control, Pino \textit{et al.} \cite{pino2022} propose a comparison of genetic programming (GP), Bayesian optimization (BO) and Lipschitz global optimization (LIPO) against DPPG on different cases, including the viscous Burgers equation and the 2D Turek problem. It was shown that GP and DDPG performed better than BO and LIPO, although DDPG had a better sample efficiency as well as a lower learning variance than GP. For optimization, the introductory PBO paper \cite{pbo} compares the latter with the standard CMA-ES algorithm, showing that PBO performs at least as well as CMA-ES.} \red{The pros and cons of DRL and canonical adjoint methods in the context of optimization problems are also discussed in~\cite{ghraieb2020}}. \blue{Obviously, additional efforts from the community should be provided in the direction of broader and more systematic comparison with existing control techniques.}

\section{Conclusion}

In the present review, the contributions of the last six years in the field of deep reinforcement learning applied to fluid mechanics problems were presented. The type of application, its complexity, the choice of control methods as well as their associated technical choices were analyzed and compared across the different contributions. Several trends and general rules of thumb currently in use in the domain were pointed out, while unusual choices and techniques were highlighted. This systematic work aims at providing a general frame of the existing usages and techniques to the researchers working in the domain, but also to help newcomers identify standard approaches and state-of-the-art performance level in the field of \blue{deep reinforcement learning}-based control for fluid dynamics.

Overall, impressive performances were observed in multiple complex control tasks. Yet, a large amount of technical questions remain unanswered, and serious efforts remain to be provided by the community in order to efficiently tackle cases of industrial-level complexity within reasonable time. In the pursue of this goal, the access to efficient \blue{computational fluid dynamics} solvers and to large computational resources remains an issue to many teams. In this perspective, the ability to successfully transfer agents from numerical to experimental environments remains to be explored more thoroughly, as the literature dealing with the coupling of \blue{deep reinforcement learning} with experimental configurations remains, to this day, extremely scarce. It makes no doubt that the upcoming years will see the mastering of these obstacles, supported by the constant progress made in the \blue{deep reinforcement learning} field and driven by the numerous industrial challenges that could benefit from it.

\appendix

\section*{Acknowledgements} 
This work is supported by the Carnot M.I.N.E.S. Institute through the M.I.N.D.S. project.

\bibliographystyle{unsrt}
\bibliography{bib}

\end{document}